\documentclass[english,a4paper]{article}
\usepackage{dcolumn}
\usepackage{bm}
\usepackage{graphicx}
\usepackage{amsmath}
\usepackage{amsfonts}
\usepackage{amssymb}
\usepackage{amsthm}
\usepackage{amstext}
\usepackage{amsbsy}
\usepackage{amsopn}
\usepackage{amscd}
\usepackage{amsxtra}

\def\sinc{\textrm{sinc}}

\def\openone{\leavevmode\hbox{\small1\kern-3.3pt\normalsize1}}

\usepackage[usenames,dvipsnames]{color}

\begin{document}

\title{Robust control of an ensemble of springs: Application to Ion Cyclotron Resonance and two-level Quantum Systems}
\author{V. Martikyan, A. Devra, D. Gu\'ery-Odelin\footnote{Laboratoire de Collisions Agr\'egats R\'eactivit\'e, Universit\'e Paul Sabatier, 118 Route de Narbonne, 31062 Toulouse Cedex 4, France}, S. J. Glaser\footnote{Department of Chemistry, Technical University of Munich, Lichtenbergstrasse 4, 85747 Garching, Germany}, D. Sugny\footnote{Laboratoire Interdisciplinaire Carnot de
Bourgogne (ICB), UMR 6303 CNRS-Universit\'e Bourgogne-Franche Comt\'e, 9 Av. A.
Savary, BP 47 870, F-21078 Dijon Cedex, France, dominique.sugny@u-bourgogne.fr}}

\maketitle

\begin{abstract}
We study the simultaneous control of an ensemble of springs with different frequencies by means of adiabatic, shortcut to adiabaticity and optimal processes. The linearity of the system allows us to derive analytical expressions for the control fields and the time evolution of the dynamics. We discuss the relative advantages of the different solutions. These results are applied in two different examples. For Ion Cyclotron Resonance, we show how to optimally control ions by means of electric field. Using a mapping between spins and springs, we derive analytical shortcut protocols to realize robust and selective excitations of two-level quantum systems.
\end{abstract}

\section{Introduction}\label{sec1}
Control processes are a key factor in many technological developments at macroscopic or microscopic scale~\cite{bryson,bressan,glaserreview,schattler}. Approaches for control design can be open-loop or closed-loop.
The second option, which is generally the most efficient, may suffer from the nature and the accuracy of the measurements required by the feedback process. These obstacles have led to the development of open-loop control techniques, which are for instance crucial in quantum control where the measurement may modify the state of the system~\cite{glaserreview,brifreview,RMP19,dongreview,altafinireview}. Different methods have been developed extending from Adiabatic processes~\cite{adiabaticreview,stirapRMP} and Optimal Control Theory (OCT)~\cite{glaserreview,bonnardbook,jurdjevicbook,pont,garon} to, more recently, Shortcut To Adiabaticity (STA)
protocols~\cite{reviewSTA1,reviewSTA2,reviewSTA3,reviewSTA4}.  In view of experimental applications, a major limitation of open-loop techniques concerns the accuracy of the modeling. This limitation can be overcome by taking into account robustness constraints in control design~\cite{glaserreview}. In this setting, adiabatic pulses are very robust but at the price of high intensity and long control duration, which can lead to undesirable effects. The original motivation of STA protocols is to speed up adiabatic control of dynamical systems, while preserving as much as possible its efficiency and robustness. Optimal process has the key advantage to minimize or maximize a specific functional, which can depend on the state of the system and on the control field. For improving robustness of non-adiabatic control pulses, a standard scenario consists in controlling an ensemble of systems which differ by the values of one or several constant parameters~\cite{liens1,liens2}. This approach has
been widely explored in quantum control, mainly by OCT~\cite{kozbar2012,kozbar2004,skinner2012,vandamme2017}, but also by STA~\cite{reviewSTA4,STAnjp,daemsprl,vandamme}. However, due to the intrinsic nonlinearity of controlled quantum dynamics, numerical algorithms are generally
used to find the control fields~\cite{grape,reichkrotov,gross,daemsprl,vandamme}. This aspect is simplified in linear systems for which formal analytical solutions can be derived even for high-dimensional dynamical processes~\cite{bryson,brockett,liberzon,Lithesis,rouchon}. In this direction, a systematic comparison between OCT and STA protocols has been recently made in a simple linear system~\cite{vartan}. Controlling linear dynamics can also be relevant in a nonlinear setting as shown recently in~\cite{li:2017}. In this work, a mapping between spins and springs allows one to design analytical and efficient broadband pulses for spin dynamics from the optimal control of an ensemble of springs.

We propose in this paper to make a general analysis of the control of an inhomogenous ensemble of linear systems by adiabatic, OCT and STA protocols. As a case study, we consider an ensemble of springs with different frequencies. Adiabatic processes are realized by means of chirped excitation pulses. Mathematical results have been established in the optimal control of such systems in ~\cite{Lithesis,li:2011,lilin}. In a completely different context, STA solutions have been also derived~\cite{guery:2014}. On the basis of these different results, we explore in this work different directions. We first show rigorously that, in the case of a continuous set of frequencies, the control field is unique for a fixed control time. In this ideal limit, we deduce that optimal and STA solutions are identical. Differences occur for a finite number of springs. Specific constraints on the control field or on the efficiency of the control process can then be taken into account. We show how these general methods can be applied in some examples and we discuss the relative advantages and flexibility of the different approaches. Finally, two concrete systems illustrate this general study. We first consider the optimal control of ions by means of electric field in Fourier-Transform Ion Cyclotron Resonance Mass Spectrometry (ICR). This technique uses a mass spectrometer based on cyclotron frequency of ions in a fixed magnetic field~\cite{bodenhausen:2016}. Ions are excited at their resonant cyclotron frequencies to a larger cyclotron radius by an oscillating electric field orthogonal to the magnetic field. Using a Rotating Wave Approximation, we show that the control process can be described by the one of a spring ensemble. The efficiency of optimal control protocols for ion excitation in a realistic setup is then highlighted. The second example is based on the nonlinear control of spins. We generalize to STA protocols the results established in~\cite{li:2017} for optimal solutions. We derive robust or selective analytical shortcut pulses for controlling an ensemble of two-level quantum systems.

The paper is organized as follows. We present the model system in Sec.~\ref{sec2} and some mathematical results about the control of a spring ensemble. Section~\ref{sec3} is dedicated to adiabatic control. The solutions derived by STA and optimal techniques are respectively presented in Sec.~\ref{sec4} and \ref{sec5}. A comparison is made and the respective advantages of the two methods are discussed. Section~\ref{sec6} focuses on the application of optimal control to ICR in order to manipulate ion trajectory. The control of spin systems by STA protocols is the subject of Sec.~\ref{controltwolevel}. Conclusion and prospective views are given in Sec.~\ref{sec7}. Technical details are reported in Appendices~\ref{appa} and \ref{appb}.
\section{The model system and mathematical results}\label{sec2}
We study the control of an ensemble of springs whose dynamics are governed by the following differential equations:
$$
\begin{pmatrix} \dot{x}_\omega \cr \dot{y}_\omega\end{pmatrix}=
\begin{pmatrix} 0 & -\omega \cr \omega & 0\end{pmatrix}
\begin{pmatrix} x_\omega \cr y_\omega\end{pmatrix}+\begin{pmatrix} u \cr 0\end{pmatrix},
$$
where $x_\omega(t)$ and $y_\omega(t)$ denote respectively the velocity and position at time $t$ of the spring of frequency $\omega$. The system is subjected to an external driving $u(t)$. We consider in this paper one control field, but the same analysis could be made for two fields along the $x$- and $y$- directions. The goal of the control is
to simultaneously steer the system from $(x_\omega(0),y_\omega(0))$ to $(x_\omega(t_f),y_\omega(t_f))$ at time $t_f$ for a continuous set of frequencies
$\omega\in [\omega_{min},\omega_{max}]$. The ensemble controllability for a continuum of Harmonic oscillators has been shown in~\cite{li:2011} if two control parameters are available. Only symmetric states of the form $x_\omega=x_{-\omega}$ and $y_\omega=-y_{-\omega}$ can be reached if only one field (in the $x$- direction) is available and the frequency range is symmetric about the origin.

As an illustrative control example, we consider as initial and final states the points $(0,0)$ and $(1,0)$ for any frequency $\omega$. By construction, we can restrict the study to positive frequencies since the target state fulfills the symmetry constraint. Note that frequency-dependent target states will be considered through the paper. If we introduce the complex coordinates $z_\omega=x_\omega+iy_\omega$, the dynamical system transforms into:
\begin{equation}\label{eqdiff}
\dot{z}_\omega=i\omega z_\omega+u.
\end{equation}
An explicit solution of Eq.~\eqref{eqdiff} is given by:
$$
z_\omega(t)=e^{i\omega t}z_\omega(0)+\int_0^t e^{i\omega (t-\tau)}u(\tau)d\tau.
$$
Since $z_\omega(0)=(0,0)$ and $z_\omega(t_f)=(1,0)$, we deduce that:
\begin{equation}\label{eqmath1}
e^{-i\omega t_f}=\int_0^{t_f}e^{-i\omega \tau}u(\tau)d\tau,
\end{equation}
for $\omega\in [\omega_{min},\omega_{max}]$.

Under some hypotheses, we show below the existence and the uniqueness of the control solution of Eq.~\eqref{eqmath1} for a continuous set of frequencies. A different proof was given in~\cite{li:2011}. We assume that $u\in L^2([0,t_f])$, i.e. $u$ is a square-integrable function with a compact support included in the interval $[0,t_f]$, $u$ is zero outside of this interval. Its Fourier transform $\hat{u}$ is an analytic function which is known over the interval $[\omega_{min},\omega_{max}]$. Since the zeros of a nonzero analytic function are isolated, we deduce
that there is at most one solution to Eq.~\eqref{eqmath1}. Indeed, if we consider two solutions $u_1$ and $u_2$ to Eq.~\eqref{eqmath1} then $\hat{u}_1-\hat{u}_2$ is
zero over $[\omega_{min},\omega_{max}]$, which contradicts the previous result. The map $\mathcal{F}$ defined by:
\begin{eqnarray*}
& & L^2([0,t_f])\to L^2([\omega_{min},\omega_{max}]) \\
& & u\mapsto \hat{u}|_{[\omega_{min},\omega_{max}]}
\end{eqnarray*}
is thus injective. The surjectivity of $\mathcal{F}$ can be described from the Paley-Wiener theorem which states the following property. The function $\hat{u}$ fulfills the condition:
$$
|\hat{u}(\omega)|\leq Ce^{t_f|\omega|},
$$
where $C>0$, if and only if there exists $u\in L^2([0,t_f])$ such that:
$$
\hat{u}(\omega)=\int_0^{t_f}e^{-i\omega \tau}u(\tau)d\tau,
$$
and we can choose $C=\int_0^{t_f}|u(\tau)|d\tau$. Satisfying the conditions of this theorem by a judicious choice of target states ensures the existence of a solution to Eq.~\eqref{eqmath1}. In the example under study, this condition is fulfilled since $|\hat{u}(\omega)|=|e^{-i\omega t_f}|=1$.

To summarize, these results establish the existence and uniqueness of an ideal mathematical control field $u(t)$ for a continuous set of frequencies. However, for practical applications, it is more interesting to consider a finite set and to take into account additional constraints on the control field. This idea will be developed for OCT and STA procedures in Sec.~\ref{sec4} and \ref{sec5} where the set of frequencies will be discretized. Note that the two fields converge towards the same solution when the discretization step goes to 0.
\section{Adiabatic control}\label{sec3}
\begin{figure}[!h]
\includegraphics[scale=0.6]{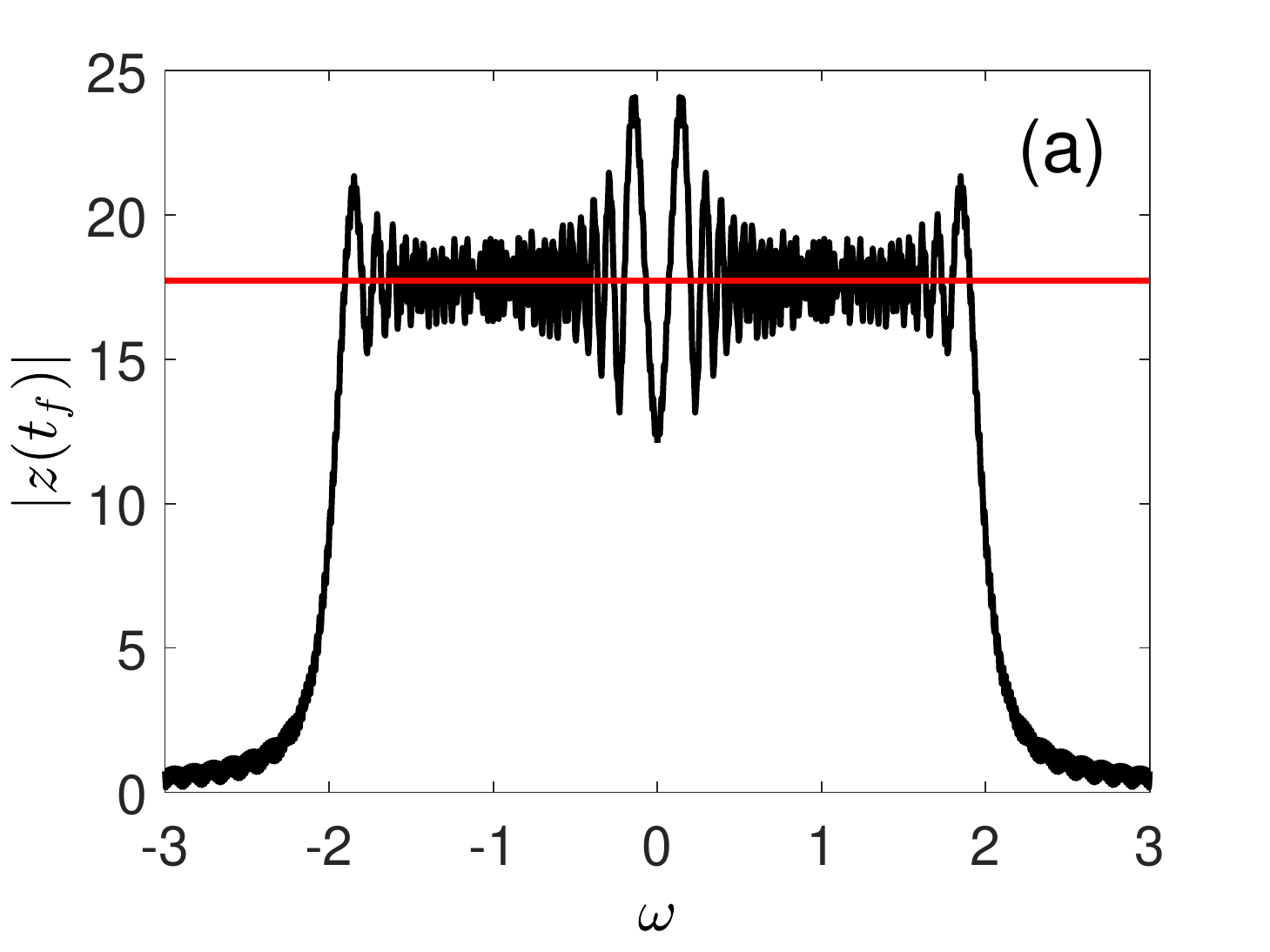}
\includegraphics[scale=0.6]{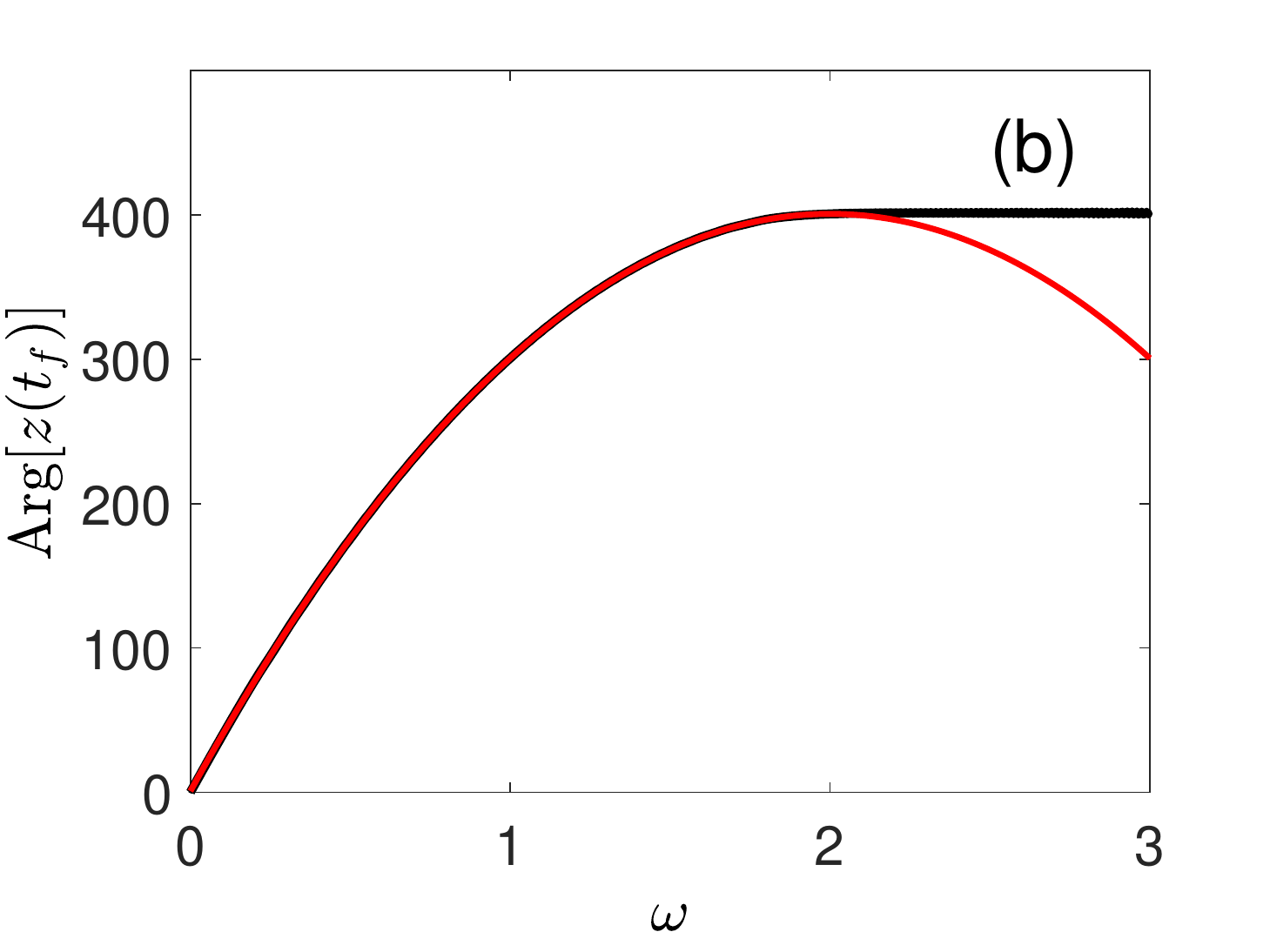}
   \caption{(Color online) Evolution as a function of $\omega$ of the radius (panel (a)) and phase (panel (b)) of an ensemble of springs, with $\omega\in [-3,3]$.
      The parameters of the adiabatic control field $u(t)$ are set to $u_0=1$, $t_f=400$, $\omega_i=0$, $\omega_f=2$ and $s=\frac{\omega_f-\omega_i}{t_f}$.
      The solid red (dark gray) lines correspond to the stationary phase approximation. Note that $|z(t_f)|$ and $\textrm{Arg}[z(t_f)]$ are respectively even and odd functions of $\omega$. Only the positive frequencies are plotted for the argument of $z(t_f)$. The different quantities are dimensionless.}
        \label{fig1}
  \end{figure}

This section is aimed at deriving an adiabatic protocol for controlling spring ensemble. This process is used below as a reference to evaluate the efficiency of OCT and STA techniques. We consider an adiabatic solution with a chirped frequency to control the spring radius. The chirp excitation pulse can be expressed as:
$$
u(t)=u_0\cos [\omega_{i} t+\frac{s t^2}{2}],
$$
where $u_0$ is the pulse amplitude, $\omega_{i}$ the initial frequency and $s$ the sweep rate. We first recall the stationary phase approximation which is used
to approximate the time evolution of the system. We consider the following integral:
$$
\hat{h}(\omega)=\int_{-\infty}^{+\infty}h(t)e^{i\phi(t)}dt,
$$
where $\phi$ is a smooth function, which is assumed to be rapidly varying with respect to $h$. A stationary point $t_0$
satisfies $\phi^{(1)}(t_0)=0$, where $\phi^{(n)}$ denotes the $n$th time derivative of $\phi$. Using a Taylor expansion around $t=t_0$, we get:
$$
\phi(t)=\phi(t_0)+(t-t_0) \phi^{(1)}(t_0)+\frac{(t-t_0)^2}{2}\phi^{(2)}(t_0)+\cdots
$$
We deduce that:
\begin{eqnarray*}
\hat{h}(\omega)&\simeq & h(t_0)e^{i\phi(t_0)}\int_{-\infty}^{+\infty}e^{i\frac{\xi^2}{2}\phi^{(2)}(t_0)}d\xi \\
 &\simeq &\sqrt{\frac{2\pi}{\phi^{(2)}(t_0)}}h(t_0)e^{i(\phi(t_0)+\frac{\pi}{4})}.
\end{eqnarray*}
For a chirp excitation, the phase $\phi(t)$ is defined by $\phi(t)=\omega_i t+\frac{s t^2}{2}$. The instantaneous frequency $\omega(t)$ can be expressed as:
$$
\omega(t)=\phi^{(1)}(t)=\omega_i+st,
$$
where $s=\omega^{(1)}(t)$. For a linear evolution of $\omega(t)$ between $\omega_i$ and $\omega_f$, the rate $s$ is given by $s=(\omega_f-\omega_i)/t_f$. We deduce that the Fourier transform of the control field is given by:
\begin{eqnarray*}
\hat{u}(\omega)&=& \int_0^{t_f} u(t)e^{-i\omega t}dt\\
&=& \frac{u_0}{2}\int_0^{t_f}[e^{i(\omega_i t+\frac{s t^2}{2}-\omega t)}+e^{-i(\omega_i t+\frac{s t^2}{2}+\omega t)}]dt.
\end{eqnarray*}
We denote by $\phi_1$ and $\phi_2$ the arguments of the two exponential terms. It is straightforward to verify that $\phi_1^{(1)}(t)=0$ for
$t=t_1^{(\omega)}=\frac{\omega-\omega_i}{s}$ and that $\phi_2^{(1)}(t)=0$ for $t=t_2^{(\omega)}=\frac{-\omega-\omega_i}{s}$. We neglect the
second contribution since $t_2^{(\omega)}<0$. If $t_1^{(\omega)}$ is not too close to 0 and $t_f$, we can consider that the
integral is defined from $-\infty$ to $+\infty$. We finally arrive at:
$$
\hat{u}(\omega)=u_0\sqrt{\frac{\pi}{2s}}e^{i(\frac{\pi}{4}+\phi_1(t_1^{(\omega)}))}.
$$
The phase spectrum $\phi(\omega)=\frac{\pi}{4}+\phi_1(t_1^{(\omega)})$ can be written as:
$$
\phi(\omega)=\frac{\pi}{4}-\frac{(\omega-\omega_i)^2}{2s}.
$$
Coming back to the original control problem, we obtain:
\begin{eqnarray}\label{eqzadia}
z_\omega (t_f)&=& e^{i\omega t_f}\int_0^{t_f}e^{-i\omega \tau}u(\tau)d\tau \nonumber \\
& \simeq & e^{i\omega t_f}u_0\sqrt{\frac{\pi}{2s}}e^{i(\frac{\pi}{4}-\frac{(\omega-\omega_i)^2}{2s})}.
\end{eqnarray}
After the adiabatic excitation, all the springs have almost the same radius, $|z_\omega(t_f)|$, but a different phase $\textrm{Arg}[z_\omega(t_f)]$, which can be expressed as:
\begin{equation}\label{eqargadia}
\textrm{Arg}[z_\omega(t_f)]=\omega t_f+\frac{\pi}{4}-\frac{(\omega-\omega_i)^2}{2s}.
\end{equation}
As can be seen in Eq.~\eqref{eqargadia}, this phase is not constant and varies quadratically with the frequency $\omega$. The radius which can be expressed as:
$$
|z_\omega(t_f)|=u_0\sqrt{\frac{\pi}{2s}},
$$
can be fixed by adjusting either the amplitude of the pulse, $u_0$, or the sweeping rate $s$. As shown in Appendix~\ref{appa}, the time evolution of the control process can be
exactly derived by using the Erfi function. A numerical example is given in Fig.~\ref{fig1}, showing the accuracy of the adiabatic approximation for a long control time $t_f$ in the range of excited springs. The main problem with this approach is its lack of flexibility since only a specific family of target states can be reached.
\section{Shortcut to Adiabaticity protocols}\label{sec4}
STA protocols correspond to fast routes between initial and ﬁnal states that are connected through a slow (adiabatic) time evolution when a control parameter is changed in time. It is thus natural to derive shortcut procedures in this control problem. STA methods generally exploit the algebraic structure of quantum mechanics~\cite{reviewSTA1,reviewSTA2,reviewSTA3,reviewSTA4}. Using inverse engineering, STA has been recently extended to statistical physics and classical mechanics. In this case, the trajectory is first extrapolated from the required boundary conditions, the shape of the control ﬁeld being deduced in a second step.
We propose in this section a general STA protocol based on a motion planning approach, known in control theory as Brunovki form~\cite{brockett,rouchon}.
We consider here a simple case in which only a discrete set of frequencies is considered and the target state is the same for all the springs. We adapt a method introduced in Ref.~\cite{guery:2014}. A general derivation for any finite-dimensional linear control system is given in Appendix~\ref{appb}. Moreover, this general approach allows us to design STA trajectory for any reachable target state, as shown in Sec.~\ref{controltwolevel}.

To clarify the construction of the control field, we first consider the case of two frequencies $\omega_1$ and $\omega_2$. We introduce an auxiliary function $g(t)$
which defines the control field:
$$
u(t)=g^{(4)}(t)+(\omega_1^2+\omega_2^2)g^{(2)}(t)+\omega_1^2\omega_2^2g(t).
$$
We show below how to determine boundary conditions on the $g$- function and its derivatives so that to realize the control process for the two springs at frequencies $\omega_1$
and $\omega_2$. The $n$th derivative of $g$ is denoted $g^{(n)}$. Assuming that $g$ obeys the following boundary conditions:
$$
g(0)=g(t_f)=g^{(1)}(0)=g^{(1)}(t_f)=g^{(2)}(0)=g^{(2)}(t_f)=0,
$$
and
$$
g^{(3)}(0)=0,~g^{(3)}(t_f)=1,
$$
an integration by parts leads to:
\begin{equation}\label{eqg2}
\int_0^{t_f}e^{-i\omega\tau}u(\tau)d\tau=e^{-i\omega t_f}+(\omega^2-\omega_1^2)(\omega^2-\omega_2^2)G(t_f),
\end{equation}
with $G(t)=\int_0^te^{-i\omega\tau}g(\tau)d\tau$. The target state is thus reached exactly for the two frequencies $\omega_1$ and $\omega_2$. For the other frequencies, the distance $d_\omega$ to the target state $(1,0)$, defined by $d_\omega=\sqrt{(x_\omega(t_f)-1)^2+y_\omega^2(t_f)}$, is given as the modulus of the second term in the right-hand side of Eq.~\eqref{eqg2}: $d_\omega=|(\omega^2-\omega_1^2)(\omega^2-\omega_2^2)G(t_f)|$. Many different solutions to this problem can be derived, such as polynomial functions but other families of functions can be chosen. A possible $g$ function is of the form:
\begin{equation}\label{eqg4}
g(t)=(\frac{t}{t_f})^4\frac{(-t_f)^3}{3!}(1-t/t_f)^3.
\end{equation}
It is then straightforward to generalize this computation to the case of $N$ frequencies. The boundary conditions are given by:
$$
\begin{cases}
g(0)=g(t_f)=0 \\
g^{(1)}(0)=g^{(1)}(t_f)=0 \\
\cdots \\
g^{(2N-2)}(0)=g^{(2N-2)}(t_f)=0 \\
g^{(2N-1)}(0)=0;~g^{(2N-1)}(t_f)=1
\end{cases}
$$
The control field can be expressed as:
$$
u(t)=\sum_{k=0}^{2N} g_kg^{(k)}(t).
$$
where the even coefficients $g_k$ (the odd coefficients are zero) are the ones of the characteristic polynomial of the diagonal matrix with the elements $(-\omega_1^2,-\omega_2^2,\cdots,-\omega_N^2)$. Here, as a possible $g$ function, we can choose:
\begin{equation}\label{eqgN}
g(t)=(\frac{t}{t_f})^{2N}\frac{(-t_f)^{2N-1}}{(2N-1)!}(1-t/t_f)^{2N-1}.
\end{equation}
Note that the $g$- function does not depend on the frequencies $\omega_k$. This process defines a family of control fields based on $g$. The distance to the target state can be determined directly from $g$:
$$
d_\omega=|\prod_{k=1}^N(\omega^2-\omega_k^2)G(t_f)|.
$$
A major limitation of this derivation relies on the definition of the $g$- function. It is thus difficult to impose constraints on the control field $u$ starting from the $g$- function. A number of frequencies lower than 10 has generally to be chosen to limit the maximum absolute amplitude of the field.

\begin{figure}[h!]
\centering
\includegraphics[width=0.5\textwidth]{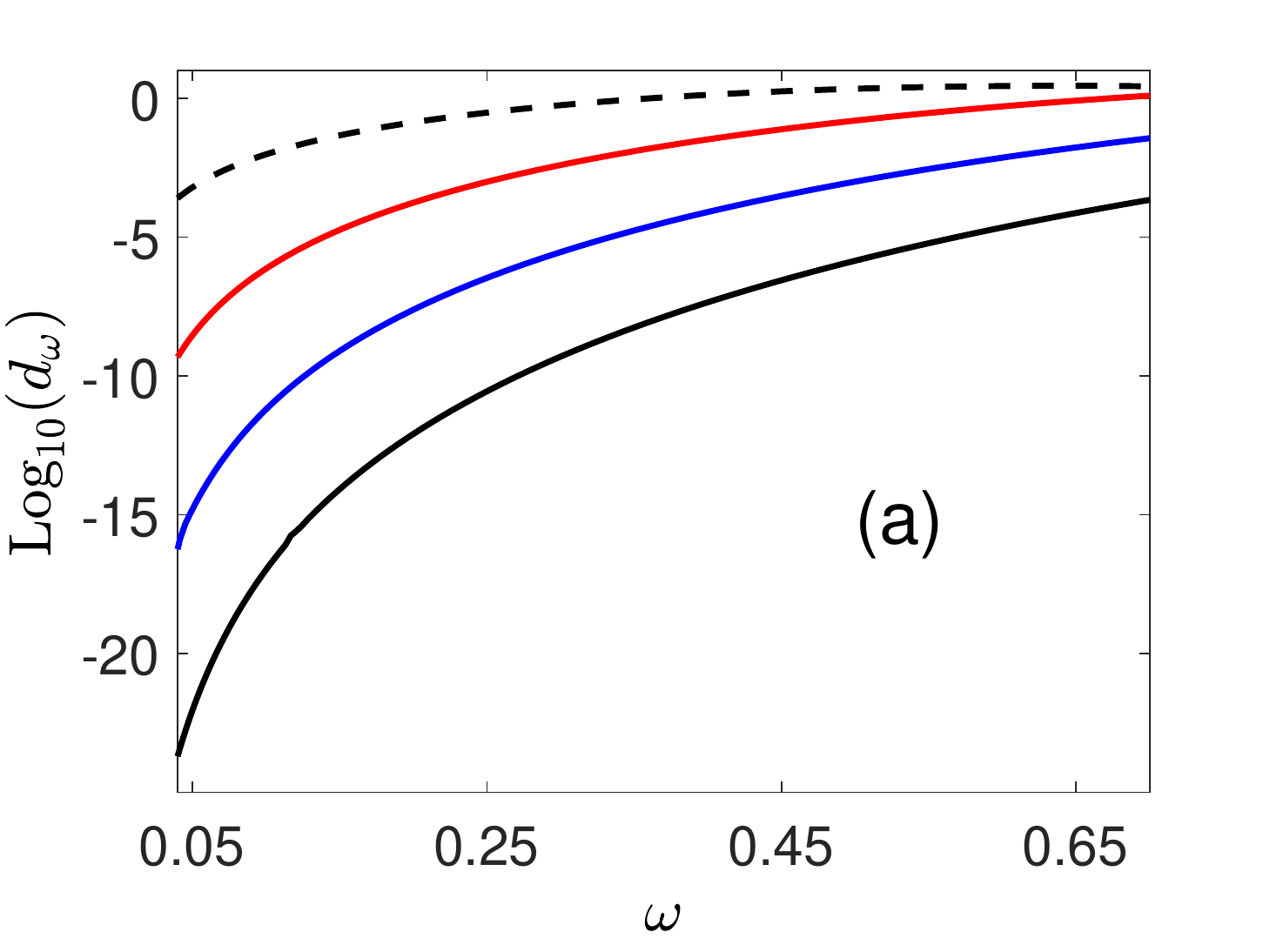}
\includegraphics[width=0.5\textwidth]{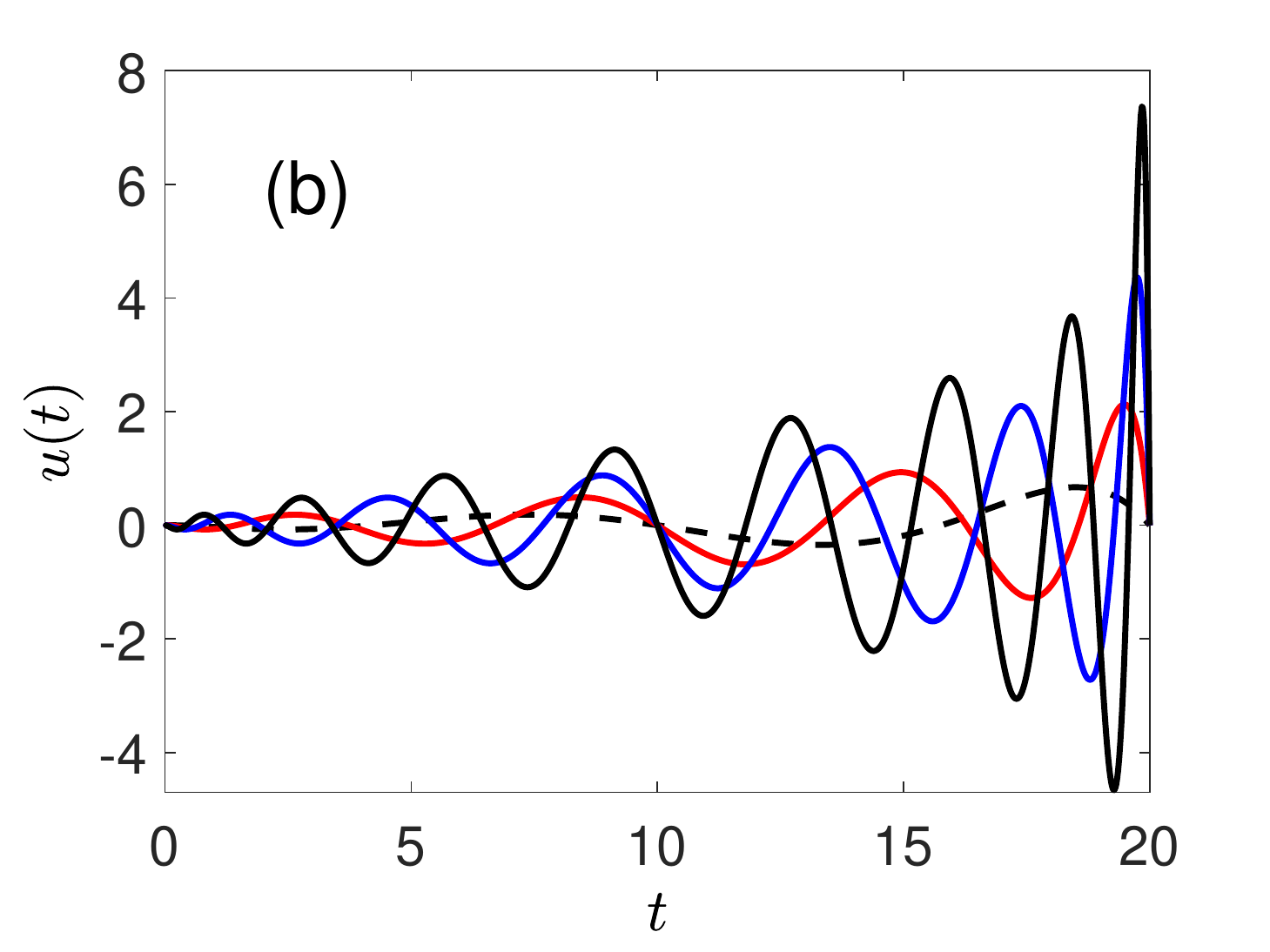}
\caption{(Color online) Ultra-high efficient STA excitation of an ensemble of springs around the frequency $\omega=0$. Panels (a) and (b) display respectively the evolution of the distance $d_\omega$ to the target state as a function of $\omega$ and the time evolution of the corresponding control field $u(t)$. Black, blue (or dark gray), red (or light gray) solid lines and dashed lines represent respectively a system with $N=8$, 6, 4 and 2 springs. Dimensionless units are used.}
\label{figsta}
\end{figure}

For $N$ springs, $4N$ boundary conditions have to be fulfilled. The minimum order of the polynomial $g$ as in Eq.~\eqref{eqgN} is therefore $4N-1$. Higher order polynomials can be derived by considering additional constraints. For instance, the initial and final values of the control field $u$ are zero if $g^{(2N)}(0)=0=g^{(2N)}(t_f)$. A solution is given by the following polynomial:
\begin{eqnarray}\label{eqgNzero}
& & g(t)=(\frac{t}{t_f})^{2N+2}\frac{(-t_f)^{2N-1}}{(2N-1)!}(1-t/t_f)^{2N-1} \\
& & \times [1+(2N+2)(1-t/t_f)].\nonumber
\end{eqnarray}
We have numerically observed that this constraint allows to limit the maximum amplitude of the pulse. Ultra-high efficient protocol around a specific frequency $\tilde{\omega}$ can be obtained if $\omega_k=\tilde{\omega}$ for any $k$. For $\tilde{\omega}=0$, the distance $d$ can be expressed as
$$
d_\omega=|\omega^{2N}\int_0^{t_f}e^{-i\omega t}g(t)dt|.
$$
Since the $g$- function does not depend on $\omega$, an upper bound to $d_\omega$ is given by $\omega^{2N}\int_0^{t_f}|g(t)|dt$. We observe that the error of the control process goes as $\omega^{2N}$ and a very good efficiency is achieved in a neighborhood of $\omega=0$ for large values of $N$. Figure~\ref{figsta} illustrates this protocol for $N=2$, 4, 6 and 8 springs. As could be expected, the error decreases as a function of $N$, while the maximum amplitude of the field increases.
\section{Optimal control theory}\label{sec5}
We focus in this section on the derivation of optimal control pulses. We consider the linear quadratic optimal control theory where the goal is to steer the system to (or close to) the target state, while minimizing the pulse energy~\cite{liberzon,rouchon}. This approach has been applied in~\cite{li:2011} to control spring ensemble for a continuous set of frequencies. The optimal solution can be expressed as an infinite expansion of prolate spheroidal wave functions. This series is then truncated to a finite set of frequencies. We propose here a different approach based on the Pontryagin Maximum Principle~\cite{pont}. We first transform the infinite dimensional control problem into a finite one by selecting a finite number of frequencies. We then apply OCT for two different cost functionals penalizing the energy of the control field. The same optimal solution as in~\cite{li:2011} is obtained by this method (Approach I) which has the advantage of being more flexible. In particular, it is straightforward to consider frequency-dependent target states.\\
\noindent\textbf{Approach I:}\\
We consider the control of a finite number $N$ of springs with frequencies $\omega_k\in [\omega_{min},\omega_{max}]$. Starting from the point $(0,0)$, the goal is to reach exactly at time $t_f$ the final states $(x_{kf},y_{kf})=z_{kf}$, where $z_k=x_k+iy_k$ is the state of the spring $k$, while minimizing the energy $E=\int_0^{t_f}u(t)^2dt$. We have:
$$
z_k(t)=\int_0^t u(\tau)e^{i\omega_k(t-\tau)}d\tau.
$$
We denote by $p_k=p_{x_k}+ip_{y_k}$ the corresponding adjoint state. The Pontryagin Hamiltonian can be expressed as:
$$
H_P=\sum_k\Re [i\omega_k z_k \bar{p}_k+p_k u]-\frac{u^2}{2},
$$
where $\Re[\cdot]$ and $\bar{[\cdot]}$ denote respectively the real part and the complex conjugate of a complex number. The dynamics of the adjoint states are governed by:
$$
\dot{p}_k=i\omega_k p_k.
$$
The optimal control is given by:
\begin{eqnarray*}
u^*(t)&=&\sum_k \Re[p_k(t)]=\sum_k \Re[p_k(0)e^{i\omega_kt}] \\
&=& \frac{1}{2}\sum_k(p_k(0)e^{i\omega_kt}+\bar{p}_k(0)e^{-i\omega_kt}).
\end{eqnarray*}
After straightforward computation, we deduce that:
\begin{eqnarray*}
\frac{2}{t_f}z_j(t_f)&=& \sum_k\exp[i\frac{(\omega_j+\omega_k)t_f}{2}]\sinc [\frac{(\omega_j-\omega_k)t_f}{2}]p_k(0)\\
& & +\exp[i\frac{(\omega_j-\omega_k)t_f}{2}]\sinc [\frac{(\omega_j+\omega_k)t_f}{2}]\bar{p}_k(0)
\end{eqnarray*}
which can be expressed in a more compact form as follows:
$$
\frac{2}{t_f}z_j(t_f)=\sum_k A_{jk}p_k(0)+B_{jk}\bar{p}_k(0),
$$
and
$$
\frac{2}{t_f}\bar{z}_j(t_f)=\sum_k \bar{B}_{jk}p_k(0)+\bar{A}_{jk}\bar{p}_k(0),
$$
where $A_{jk}=\exp[i\frac{(\omega_j+\omega_k)t_f}{2}]\sinc [\frac{(\omega_j-\omega_k)t_f}{2}]$ and $B_{jk}=\exp[i\frac{(\omega_j-\omega_k)t_f}{2}]\sinc [\frac{(\omega_j+\omega_k)t_f}{2}]$. Solving this linear system, we get the initial adjoint states and therefore the optimal control field and the optimal trajectories. Note that numerical errors appear if the linear system is close to a singular system.

\begin{figure}[h!]
\centering
\includegraphics[width=0.5\textwidth]{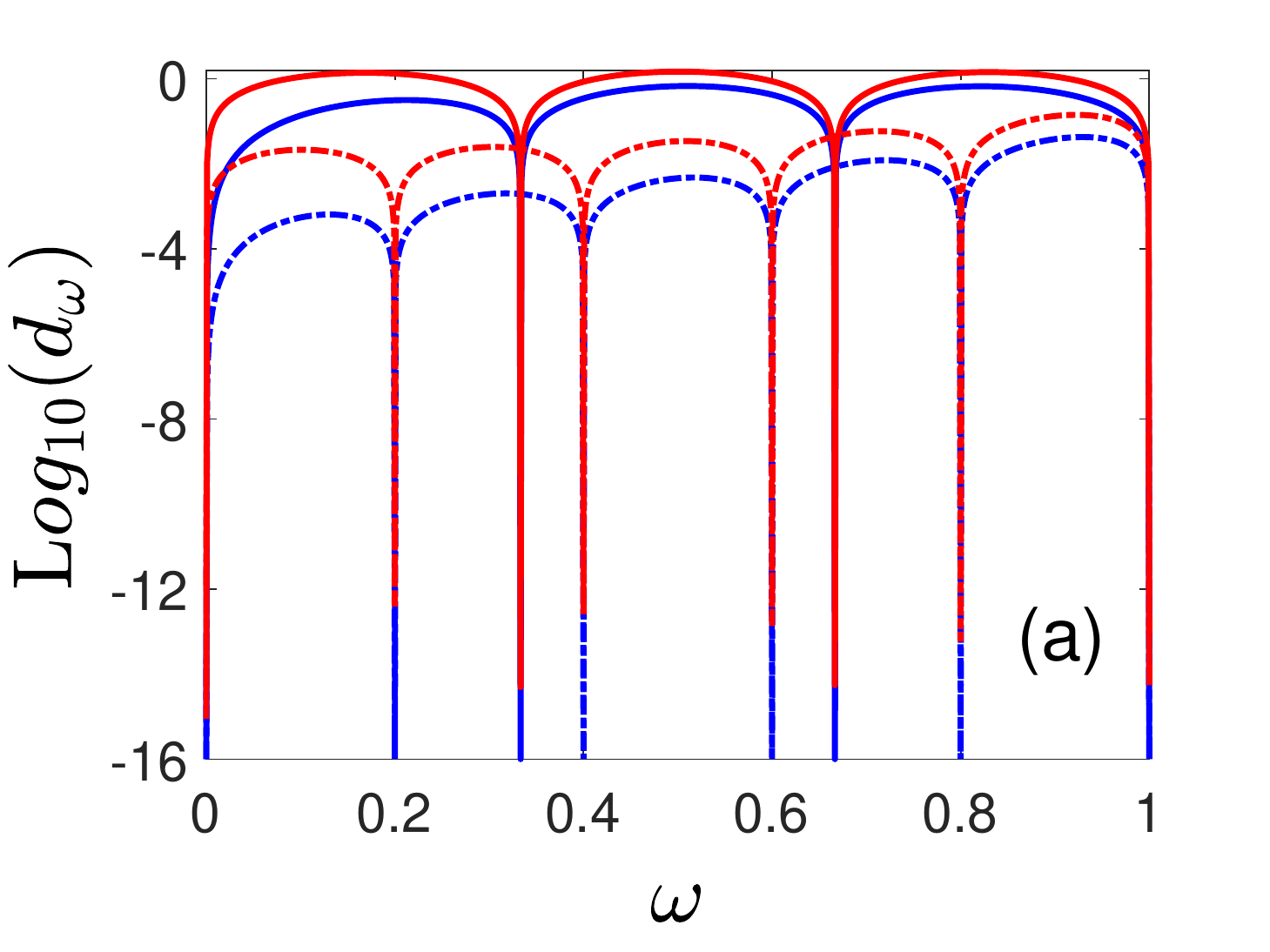}
\includegraphics[width=0.5\textwidth]{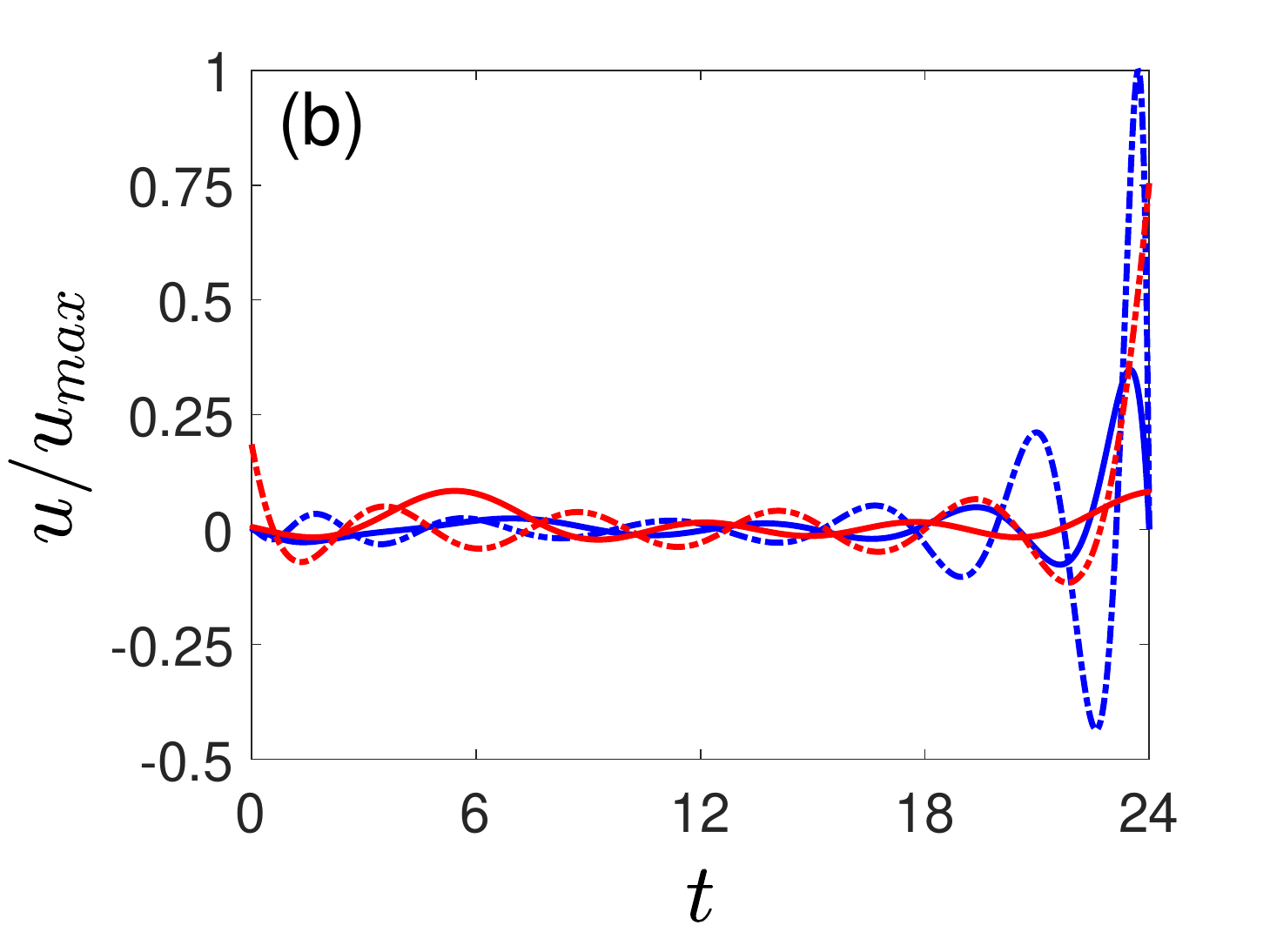}
\caption{(Color online) STA (blue or dark gray) and optimal (red or light gray) excitations of an ensemble of springs in the range of frequencies $\omega\in [0,1]$. The pulses have been computed for a regular distribution of $N=4$ (solid line) and 6 (dashed line) springs. Panels (a) and (b) display respectively the distance to the target state and the corresponding control fields. The control time is set to $t_f=24$. Dimensionless units are used.}
\label{figstaoct}
\end{figure}

A comparison of this method with STA protocols introduced in Sec.~\ref{sec4} is presented in Fig.~\ref{figstaoct} for a spring ensemble with $\omega\in [0,1]$. As above, the goal is to transfer the system from the point $(0,0)$ to $(1,0)$ in a time $t_f$. We consider two regular discretizations with $N=4$ and $6$ frequencies. The parameters of the different pulses are given in Tab.~\ref{tab1}. As could be expected, we observe a strong similarity between STA and OCT solutions. The distance to the target state is very small for points which do not belong to the grid frequency. Slightly better results are achieved with STA processes, but at the price of more energetic pulses. The target states are not exactly reached with the optimal process because the linear system used to determine the control field is close to a singular one.\\
\begin{table}[tb]
\caption{Comparison between OCT and STA pulses for controlling a spring ensemble. The control time is set to $t_f=24$. $u_{max}$ and $E$ denote respectively the maximum absolute value of the control field and the normalized energy $E=\int_0^{t_f}u(t)^2dt$.\label{tab1}}
\begin{tabular}{|c|c|c|}
\hline
 & $N=4$ & $N=6$\\
\hline
\hline
$u_{max}$ (STA) & 1.10 & 3.16\\
\hline
$u_{max}$ (OCT) & 0.27 & 2.38\\
\hline
E (STA) & 1.03 & 6.06\\
\hline
E (OCT) & 0.26 & 2.39\\
\hline
\end{tabular}
\end{table}
\noindent\textbf{Approach II}\\
We consider a second approach where the distance to the target states (for a finite set of frequencies $\omega_k$) is defined in the cost functional $\mathcal{J}$ to minimize. The cost functional $\mathcal{J}$ can be expressed as:
$$
\mathcal{J}=\sum_k \frac{1}{2}[(x_k(t_f)-x_{kf})^2+(y_k(t_f)-y_{kf})^2]+\frac{\lambda}{2}\int_0^{t_f}u^2dt,
$$
where $\lambda$ is a positive penalty factor chosen to weight the importance of the pulse energy. The Pontryagin Hamiltonian is:
$$
H_P=\sum_k\Re [i\omega_k z_k \bar{p}_k+p_k u]-\frac{\lambda u^2}{2},
$$
and the optimal control is given by:
$$
u^*=\frac{1}{\lambda}\sum_k\Re [p_k]
$$
The time evolution of $p_k$ can be expressed as:
$$
p_k(t)=p_k(0)e^{i\omega_k t}=p_k(t_f)e^{i\omega_k (t-t_f)},
$$
with the final condition:
$$
p_k(t_f)=z_{kf}-z_k(t_f).
$$
After straightforward computation, we deduce that:
\begin{eqnarray*}
\frac{2\lambda}{t_f}z_j(t_f)=& &\sum_k \exp[i\frac{(\omega_j-\omega_k)t_f}{2}]\sinc [\frac{(\omega_j-\omega_k)t_f}{2}]p_k(t_f)\\
& & +\exp[i\frac{(\omega_j+\omega_k)t_f}{2}]\sinc [\frac{(\omega_j+\omega_k)t_f}{2}]\bar{p}_k(t_f),
\end{eqnarray*}
which can be expressed as:
\begin{equation}\label{octeq2}
\frac{2\lambda}{t_f}z_j(t_f)=\sum_k C_{jk}(z_{kf}-z_k(t_f))+D_{jk}(\bar{z}_{kf}-\bar{z}_k(t_f)),
\end{equation}
%and
%$$
%\frac{2\lambda}{t_f}\bar{z}_j(t_f)=\sum_k \bar{C}_{jk}(\bar{z}_{kf}-\bar{z}_k(t_f))+\bar{D}_{jk}(z_{kf}-z_k(t_f)),
%$$
with
$$
\begin{cases}
C_{jk}=\exp[i\frac{(\omega_j-\omega_k)t_f}{2}]\sinc [\frac{(\omega_j-\omega_k)t_f}{2}] \\
D_{jk}=\exp[i\frac{(\omega_k+\omega_j)t_f}{2}]\sinc [\frac{(\omega_k+\omega_j)t_f}{2}].
\end{cases}
$$
Equation~\eqref{octeq2} and its complex conjugate give the dynamical state at time $t_f$, and thus the final adjoint state. We then obtain the control field $u(t)$. The efficiency of this second approach is shown in Sec.~\ref{sec6} for controlling ion dynamics.
\section{Ion Cyclotron Resonance}\label{sec6}
\begin{figure}[h!]
\centering
\includegraphics[width=0.5\textwidth]{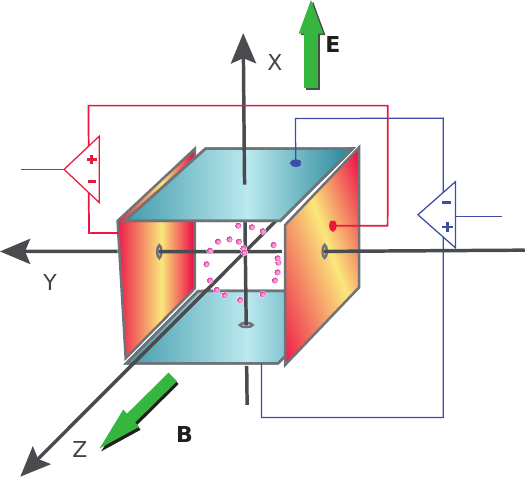}
\caption{(Color online) Schematic description of the control of ions in ICR. The pink (or black) dots represent the time evolution of the ions inside the cell. A homogeneous constant magnetic field is applied along the $z$- axis. The ion excitation is controlled by a time-dependent electric field along the $x$- direction, which is generated by a voltage difference between the blue (dark gray) plates. The position of the ion is measured by the charge induced on the red (light gray) plates.}
\label{figicr}
\end{figure}

The Fourier-Transform Ion Cyclotron Resonance (ICR) mass spectrometry is a type of mass spectrometer based on cyclotron frequency of ions in a fixed magnetic field~\cite{bodenhausen:2016,delsuc:2013,delsuc:2016,delsuc:2017}. Ions are trapped in a Penning trap, where they are excited by an electric field. After the excitation process, the ions rotate at their cyclotron frequency as a \emph{packet} of ions. The image charge induced by the ions on a pair of electrodes is detected. The Fourier transform of the resulting transient signal leads to the mass spectrum. ICR allows to access the highest resolution available in mass spectrometry. A schematic representation of the experimental setup is given in Fig.~\ref{figicr}. In this section, we propose to show how optimal control can be used to design excitation pulses in ICR. Standard processes in this domain are based on adiabatic chirped pulses. Optimal control should allow a much wider range of possibilities, such as a precise and robust control of ion radius and a linear frequency dependence of the phase. As shown in Sec.~\ref{sec4}, the phase evolves quadratically with the frequency in adiabatic control. While a general study of this process goes beyond the scope of this work, we propose to analyze a simplified version in which the rotating wave approximation (RWA) can be applied. In this setting, the robust control of ions is described by the one of a spring ensemble and the material of Sec.~\ref{sec5} can be directly used.
\subsection{The model system}
The different ions in the experimental cell are subjected to a magnetic field $\vec{B}$ along the $z$- axis and to an electric field $\vec{E}$ in the $(x,y)$- plane~\cite{bodenhausen:2016,delsuc:2013,delsuc:2016,delsuc:2017}. The dynamics are governed by the Lorentz's equation:
\begin{equation}
m_k\dot{\vec{v}}_k=q_k\vec{E}+q_k(\vec{v}_k\times \vec{B}),
\end{equation}
which can be expressed as:
\begin{equation}
\begin{cases}
\dot{v}_{x_k}=\omega_k(e_x+v_{y_k}) \\
\dot{v}_{y_k}=\omega_k(e_y-v_{x_k}),
\end{cases}
\end{equation}
with $\omega_k=\frac{q_k B}{m_k}$ and $\vec{e}=\vec{E}/B$. The frequency $\omega_k$ belongs to the interval $[\omega_{min},\omega_{max}]$. We consider now the complete control problem with the speed and the position of the different ions. The dynamics are governed
for the ion $k$ by the following differential system:
\begin{equation}
\begin{cases}
\dot{x}_k=v_{x_k} \\
\dot{y}_k=v_{y_k} \\
\dot{v}_{x_k}=\omega_k (e_x+v_{y_k}) \\
\dot{v}_{y_k}=\omega_k (e_y-v_{x_k}).
\end{cases}
\end{equation}
In practical applications, only the electric field $e_x$ along the $x$- direction is available for controlling ions, $e_y(t)=0$.
Starting from the center of the cell $(x_k=0,y_k=0)$, the goal is to reach at a fixed control time a given radius with either a constant
phase with respect to $\omega$ or with a phase varying linearly with $\omega$. In standard experiments, a chirped adiabatic excitation is used
and leads to a control of the radial coordinate but not of the phase.
\subsection{Rotating Wave Approximation}
We describe in this section the RWA which allows to simplify the control of ICR processes. Using this approximation, we show that the control of ions reduces to the control of an ensemble of springs of different frequencies. We start with the speed control which satisfies:
$$
\begin{cases}
\dot{v}_{xk}=\omega_k v_{yk}+\omega_k e_x \\
\dot{v}_{yk}=-\omega_kv_{xk}
\end{cases}
$$
In complex coordinates, we have:
$$
\dot{V}_k=-i\omega_kV_k+\omega_k e_x(t),
$$
where $V_k=v_{xk}+iv_{yk}$. We assume that $\omega_k\in [\omega_0-\delta\omega,\omega_0+\delta\omega]$ and $e_x(t)=e_0(t)\cos(\omega_0 t)$, where $\delta\omega \ll \omega_0$ and $e_0(t)$ varies slowly in time (slowly varying envelope approximation). We express the complex speed as: $V_k=\tilde{V}_ke^{-i\omega_0 t}$. We deduce that:
$$
\dot{\tilde{V}}_k=-i\Delta\omega_k\tilde{V}_k+\omega_k\frac{e_0}{2}(1+\exp(-2i\omega_0 t)),
$$
where $\Delta\omega_k=\omega_k-\omega_0$ is the detuning term. In RWA, we neglect the rapidly oscillating term $\exp(-2i\omega_0 t)$ and we arrive at:
$$
\dot{\tilde{V}}_k=-i\Delta\omega_k\tilde{V}_k+\omega_k\frac{e_0}{2}.
$$
We recover the control of an ensemble of springs by assuming that $\omega_k\simeq \omega_0$ for any ion, i.e. we replace the term $\omega_k \frac{e_0}{2}$ by $\omega_0 \frac{e_0}{2}$. An additional approximation can be made for the position of the ion. We set $X_k=\tilde{X}_ke^{-i\omega_0 t}$.
It is then straightforward to show that:
$$
\dot{\tilde{X}}_k-i\omega_0\tilde{X}_k=\tilde{V}_k(t)
$$
Since $\tilde{X}_k$ varies slowly with respect to $e^{i\omega_0 t}$, we can neglect the time derivative $\dot{\tilde{X}}_k$, which gives:
$$
\tilde{X}_k=\frac{i}{\omega_0}\tilde{V}_k(t).
$$
In this limit, we deduce that the speed control leads also to the control of the position of ions.
\subsection{Numerical results}
We illustrate the optimal control of ions with the following numerical example. We consider the Approach II presented in Sec.~\ref{sec5}.

We first compute the optimal control $u(t)$ of a spring ensemble with $\omega\in [0,200]$. The control time $t_f$ is set to 1. At this point, all the quantities are dimensionless. The target states $z_{f\omega}$ depend on the frequency and the final radius of the trajectory can be expressed as:
$$
|z_{f\omega}|=\frac{1}{2}(1+\tanh((\omega_S-\omega)\mu)),
$$
where $\mu=0.1$ and $\omega_S=100$. The target radius is of the order of 1 for $\omega<\omega_S$ and 0 for $\omega>\omega_S$. The smooth transition between the two regions can be adjusted with the parameter $\mu$. The phase of the target state is defined as:
$$
z_f(\omega)=|z_{f\omega}|\exp(i\omega \eta t_f)
$$
with $\eta=0.5$, the slope of the frequency-dependent phase. We observe numerically that a non-zero slope in a given range ($\eta\in ]0,1[$) helps limit the maximum amplitude of the pulse. The same observation was made for spin control~\cite{slope1,slope2}. The parameter $\lambda$ of the approach II, which weights the importance of the pulse energy in the cost functional, is set to $10^{-3}$. A regular discretization of 60 frequencies in the range [0,200] is taken into account in the optimization. Note that the final result does not change if a sufficient number of frequencies is used.

The control field is then expressed in physical units as follows. We define the normalized electric field $e(t)$ as:
$$
e(t)=\frac{E_0}{B_0}u(t)\cos(\omega_0 t),
$$
where $E_0=100$~V.m$^{-1}$, $B_0=10$~T and $\omega_0/(2\pi)=500$~kHZ. These values are typical of ICR spectrometers. The intensity of the electric field $E_0$ is fixed to get a radial excitation of a few centimeters. The control time is assumed to be expressed in ms, leading to a control duration of 1~ms, which is also standard in ICR. We deduce that a range of $\Delta\omega/(2\pi)=100/(2\pi)=16$~kHz is excited around the central frequency $\omega_0/(2\pi)$. Note that the RWA is justified since $\Delta\omega\ll \omega_0$.

\begin{figure}[h!]
\centering
\includegraphics[width=0.5\textwidth]{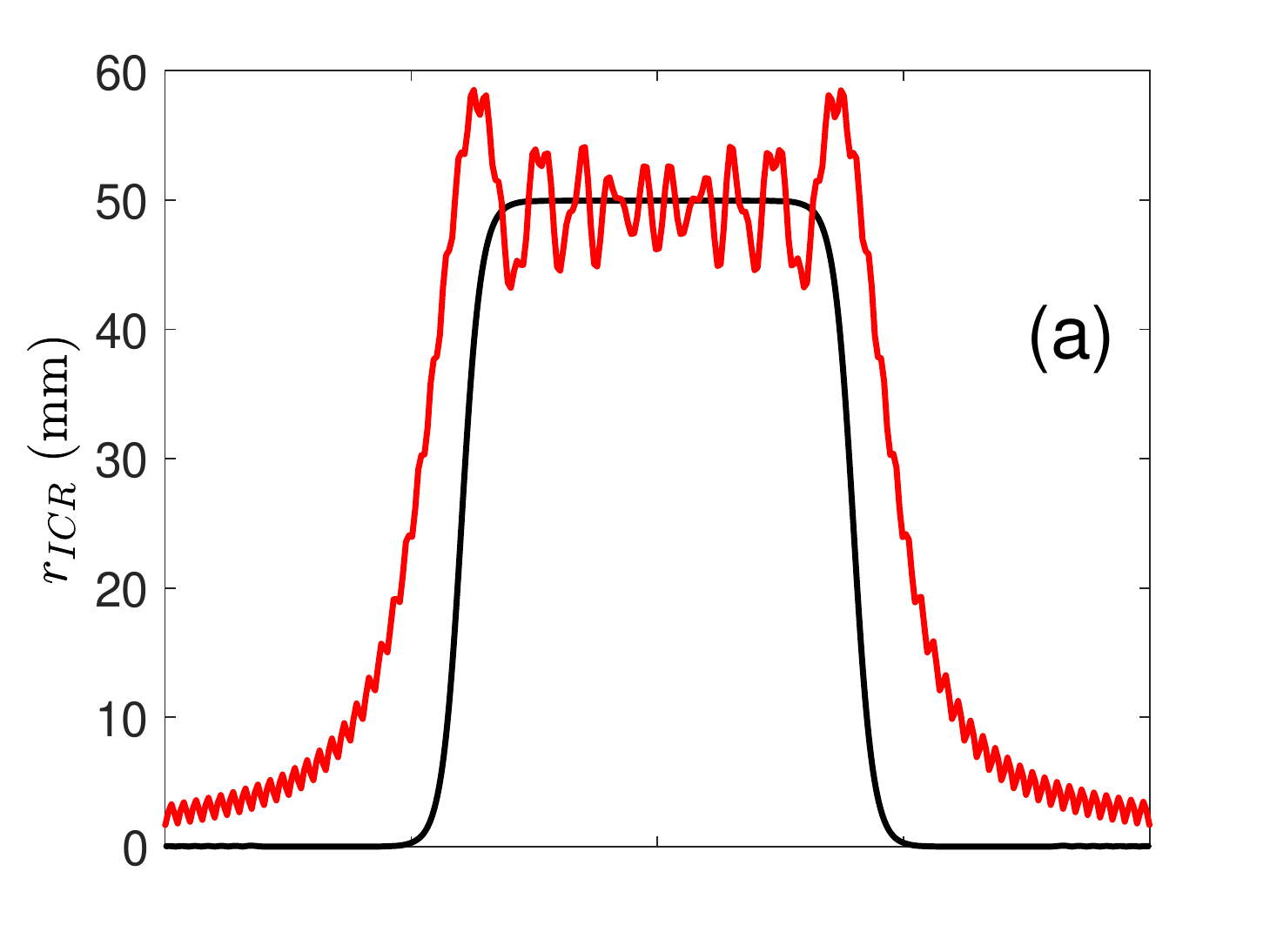}
\includegraphics[width=0.5\textwidth]{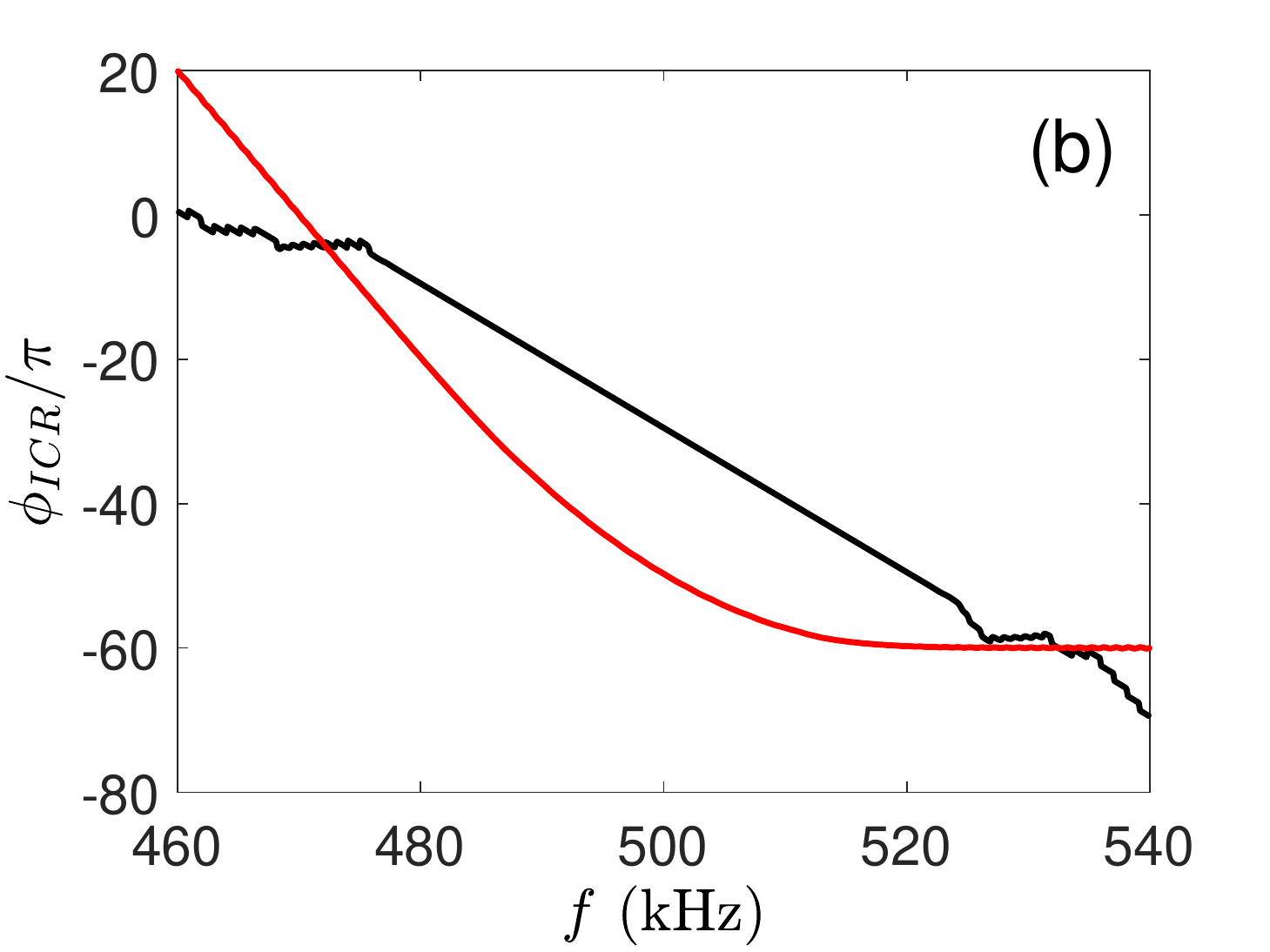}
\includegraphics[width=0.5\textwidth]{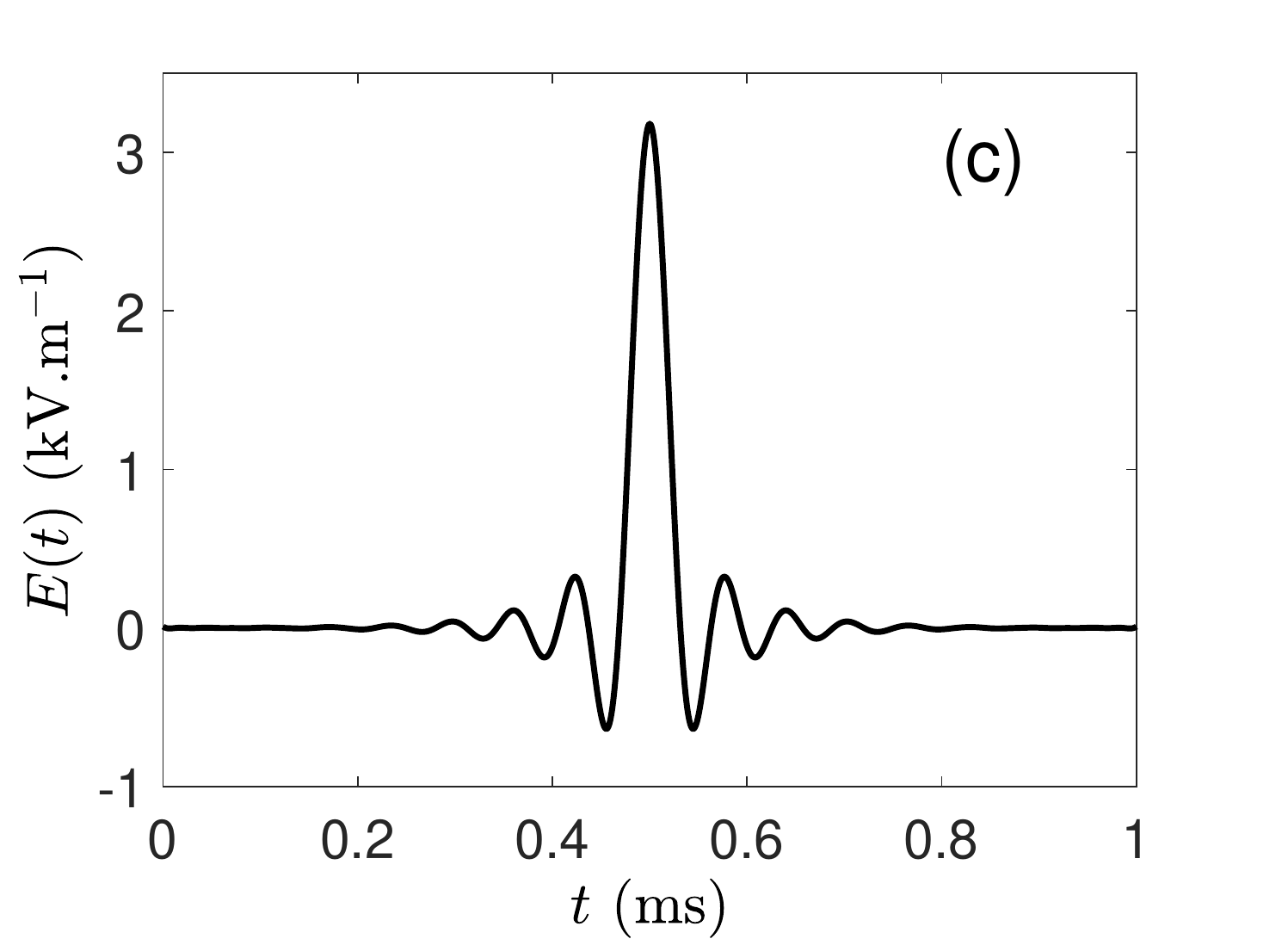}
\caption{(Color online) Excitation of an ensemble of ions by optimal (black) and adiabatic (red or light gray) pulses. Panels (a) and (b) depict respectively the final radius $r_{ICR}$ (in mm) and the final phase $\phi_{ICR}$ (in radian) of ions as a function of the frequency $f$ in the range of frequencies $[460,540]$~kHz. In panel (b), an arbitrary vertical shift has been added to the phase of the adiabatic excitation to ease the comparison. Panel (c) displays the optimal control pulse $E(t)$ with a duration of 1~ms.}
\label{figion}
\end{figure}

Numerical results are presented in Fig.~\ref{figion}. The radius and the phase of the ion are denoted $r_{ICR}$ and $\phi_{ICR}$. A comparison can be made with an adiabatic excitation, characterized by the following parameters: $\omega_{i}/(2\pi)=480$~kHz, $\omega_{f}/(2\pi)=520$~kHz, $t_f=1$~ms and an amplitude $E_0=0.625$~kV$\cdot$m$^{-1}$. The sweep rate $s$ is defined as $s=\frac{\omega_f-\omega_i}{t_f}$. We observe that the optimal control process generates a very good excitation inside the expected range of frequencies. This control procedure is directly comparable to the adiabatic process.
\section{Control of two-level quantum systems}\label{controltwolevel}
Performing fast and efficient control of two-level quantum systems represents a crucial prerequisite in different domains going from Nuclear Magnetic Resonance to quantum computing~\cite{glaserreview}. The design of robust or selective control processes has been the subject of intense progress in the last decades. Different methods based either on composite pulses~\cite{genov:2014,levitt:1986,rong:2015}, STA principles~\cite{reviewSTA4,STAnjp,daemsprl,vandamme} or OCT~\cite{glaserreview,kozbar2004,kozbar2012,vandamme2017,selective} have been developed. However, due to the nonlinearity of the dynamics, the control fields are generally determined by numerical methods. In this setting, a breakthrough idea was recently proposed in~\cite{li:2017}. Using the mapping between spin and spring and the linearization of the associated dynamics, it was shown that efficient broadband analytical pulses can be derived from the control of the linear system. This approach was illustrated with optimal control procedures. We propose in this work to show that this mapping can be extended to STA solutions. For that purpose, we consider both robust and selective control processes based on the STA approach presented in Sec.~\ref{sec4}.
\subsection{The model system}\label{lintwolevel}
We first describe the mapping between the nonlinear and linear systems. We consider the control of a two-level quantum system whose dynamics are governed in Bloch representation~\cite{levittbook,lapert:2010} by:
\begin{equation*}
\begin{cases}
\dot{x}=-\omega y+uz \\
\dot{y}=\omega x \\
\dot{z}=-ux
\end{cases}
\end{equation*}
where $(x,y,z)$ are the Bloch coordinates, with $x^2+y^2+z^2=1$, $\omega$ is the offset term and $u(t)$ the control field. Using the spherical coordinates $(\theta,\phi)$ defined by $x=\sin\theta\cos\phi$, $y=\sin\theta\sin\phi$ and $z=\cos\theta$, we arrive at:
\begin{equation*}
\begin{cases}
\dot{\theta}=u\cos\phi \\
\dot{\phi}=\omega-u\sin\phi\cot\theta.
\end{cases}
\end{equation*}
The Laurent series of the cotan function around $\theta=0$:
$$
\cot\theta=\frac{1}{\theta}-\frac{1}{3}\theta-\frac{1}{45}\theta^3+\cdots ,
$$
leads to:
\begin{equation}\label{eqtwo}
\begin{cases}
\dot{\theta}=u\cos\phi \\
\dot{\phi}=\omega-u\sin\phi (\frac{1}{\theta}-\frac{1}{3}\theta-\frac{1}{45}\theta^3).
\end{cases}
\end{equation}
The dynamical equation of the spring is:
\begin{equation*}
\begin{cases}
\dot{x}=-\omega y+u \\
\dot{y}=\omega x.
\end{cases}
\end{equation*}
Introducing the polar coordinates $(r,\Phi)$ such that $x = r\cos\Phi$ and $y = r\sin\Phi$, we get:
\begin{equation*}
\begin{cases}
\dot{r}=u\cos\Phi \\
\dot{\Phi}=\omega-\frac{u\sin\Phi}{r},
\end{cases}
\end{equation*}
which can be identified to the two-level system of Eq.~\eqref{eqtwo} at first order in $\theta$ where $\cot\theta \simeq \frac{1}{\theta}$. In this mapping, $r$ and $\Phi$ are respectively associated to $\theta$ and $\phi$. This identification can be used for a broadband excitation process in which the spin goes from the state $(x=0,y=0,z=1)$ to $(1,0,0)$ or from $(\theta=0,\phi=0)$ to $(\frac{\pi}{2},0)$. The spin inversion can be realized by combining two successive excitation protocols (with the second one in time reversed order)~\cite{li:2017}.
\subsection{Robust and selective control}
\begin{figure}[h!]
\centering
\includegraphics[width=0.5\textwidth]{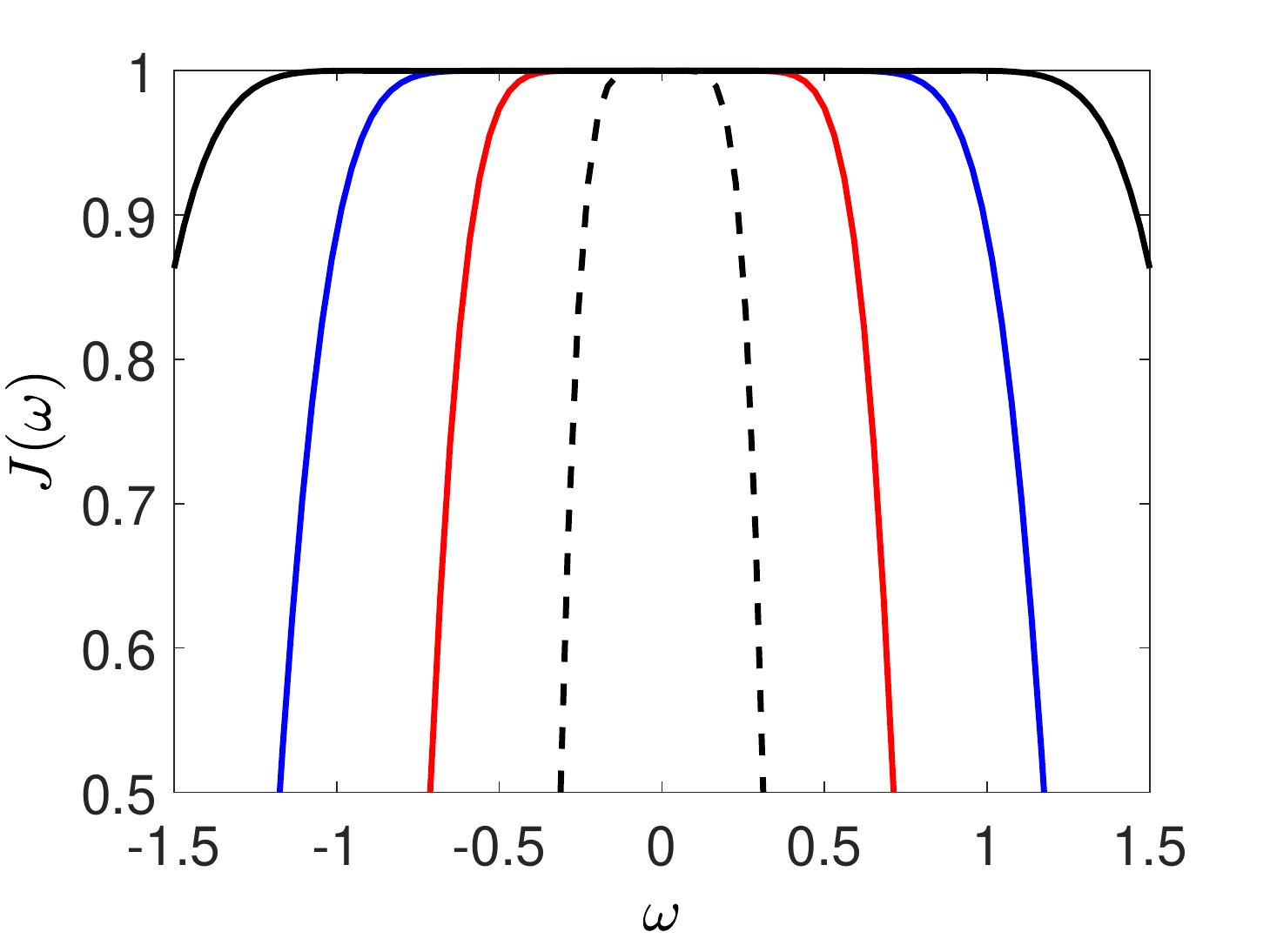}
\caption{(Color online) Robust inversion of an ensemble of two-level quantum systems with respect to the offset $\omega$ by means of spring STA protocols. The color code is the same as in Fig.~\ref{figsta}, i.e. dashed line and solid blue (or dark gray), red (or light gray) or black lines for $N=2$, 4, 6 and 8 springs, respectively. Dimensionless units are used.}
\label{figinv}
\end{figure}

\begin{figure}[h!]
\centering
\includegraphics[width=0.5\textwidth]{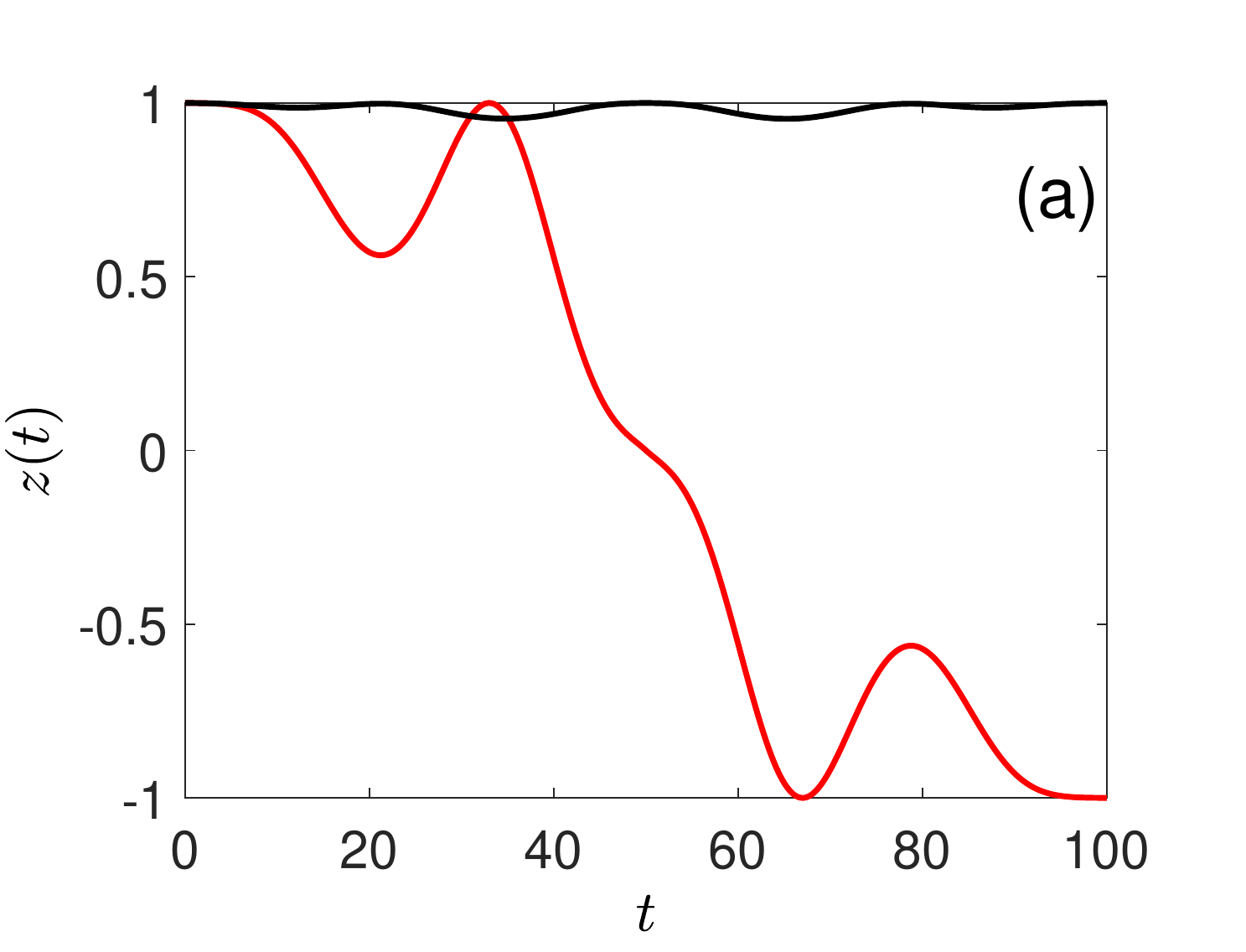}
\includegraphics[width=0.5\textwidth]{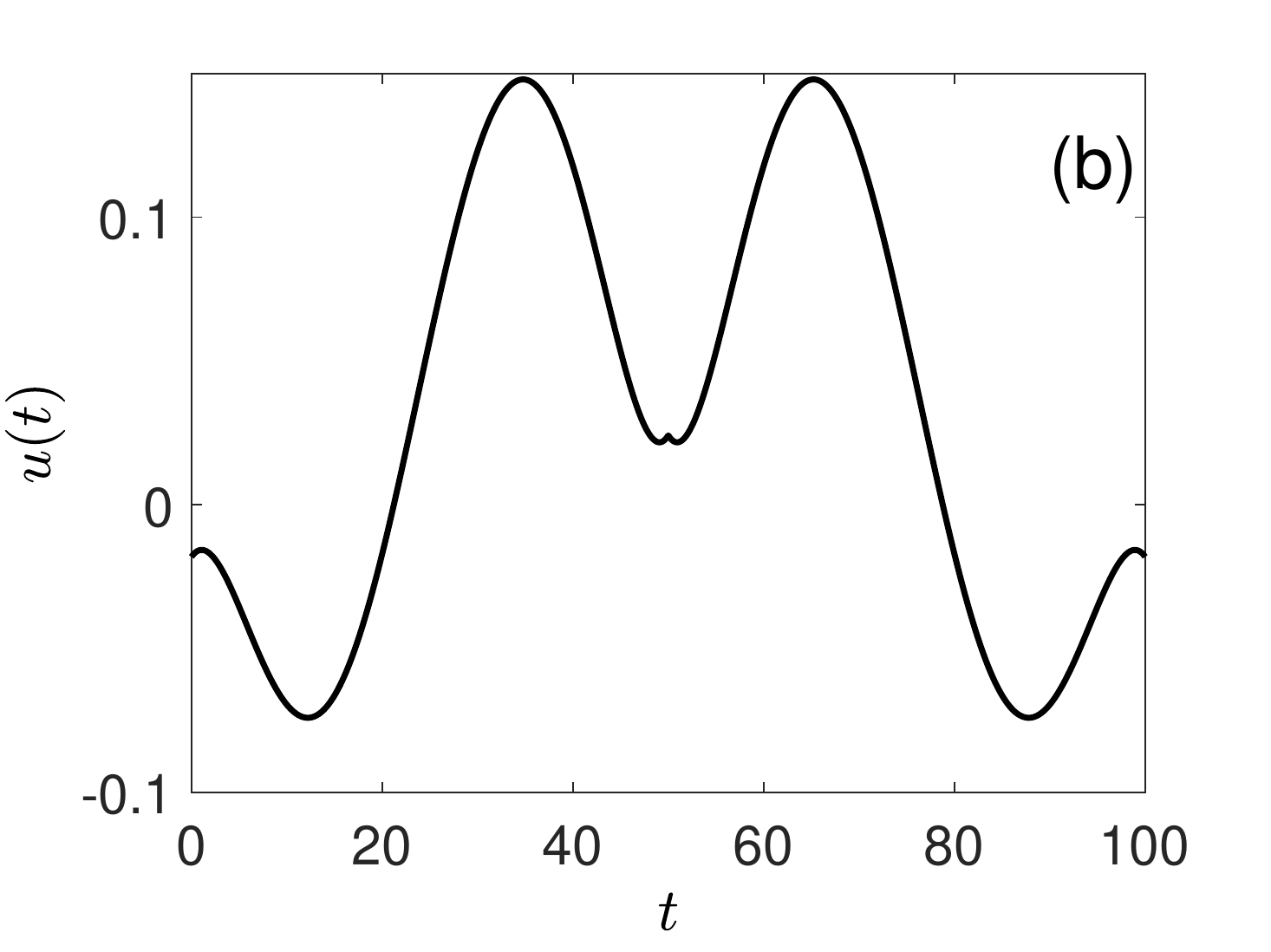}
\caption{(Color online) Selective inversion of two quantum systems of frequencies 0 and 0.5. Panel (a) represents the time evolution of the $z$- coordinate of the two systems in black and red (or dark gray) solid lines. The control field is displayed in Panel (b). Dimensionless units are used.}
\label{figsel}
\end{figure}

We illustrate the efficiency of STA control protocols derived in Sec.~\ref{sec4} on two examples, namely the robust and the selective control of two-level quantum systems. A first example is given in Fig.~\ref{figinv} for the inversion process by using control fields of Fig.~\ref{figsta}. Note that the pulses have been scaled by a factor $\frac{\pi}{2}$ since the spring goes here from $(x=0,y=0)$ to $(\frac{\pi}{2},0)$. The fidelity of the control $J(\omega)$ is defined as $J(\omega)=-z(t_f)$ for a specific offset $\omega$. A fidelity of 1 indicates that the process is perfectly realized. We observe in Fig.~\ref{figinv} the remarkable efficiency of the control protocol for a large range of frequencies.
A second example is given in Fig.~\ref{figsel} for a selective process. We consider two quantum systems of frequencies $\omega_1=0$ and $\omega_2=0.5$. The goal of the control protocol is to invert the first system, while returning for the second spin to the initial state. Using the general procedure of Appendix~\ref{appb}, a STA solution can be obtained with two springs starting from $(0,0)$ and going to the final points $(\frac{\pi}{2},0)$ and $(0,0)$. Here again, the control field is applied two times to perform the selective inversion of the spin. We observe that a long duration is needed to limit the maximum amplitude of the pulse. The second spin remains close to the origin during the control process.
These two different examples show the efficiency and the flexibility of the spin-spring mapping to derive analytical pulses able to control an ensemble of two-level quantum systems. A polynomial basis has been used, but the control field could be expanded over other basis functions, adapted to specific applications. The maximum intensity of the pulse cannot be directly fixed by this approach. It can be changed by playing with the control duration.
\section{Conclusion}\label{sec7}
We have reviewed in this study different approaches to control the dynamics of an inhomogeneous ensemble of springs. The different methods presented in this paper can be used in any linear control system. They also provide interesting alternatives to design pulses controlling two-level quantum systems.
We have shown that STA and optimal protocols may exceed the limits of adiabatic control. Any target state and control duration can be formally chosen, which can lead, e.g., to robust or selective control protocols. In order to satisfy experimental limitations on the shape of the control field, additional constraints have to be accounted for. For the two methods, only a finite set of frequencies (with a regular discretization) are considered. This aspect has not been described in this paper, but this frequency set can be optimized in a practical application to improve the efficiency of the control process. We have also discussed the relative advantages of the two methods. The efficiency of the derived control fields is comparable. STA allows to derive simple and smooth control solutions, which can be expanded in a given basis of functions. However, it is difficult to account for additional constraints on the amplitude or the energy of the pulse, requirements that can be fulfilled with OCT.

\appendix

\section{Exact derivation of adiabatic dynamics}\label{appa}
We show in this paragraph that the time evolution of adiabatic dynamics can be exactly derived. For that purpose, we need to compute integrals of the form:
$$
\mathcal{I}(\alpha,\beta)=\int_0^{t_f}\exp[i\alpha t^2+i\beta t]dt,
$$
where $\alpha$ and $\beta$ are real coefficients. We have:
$$
\mathcal{I}(\alpha,\beta)=\exp[-i\frac{\beta^2}{4\alpha}]\int_0^{t_f}\exp[\big(e^{i\pi/4}\sqrt{\alpha} (t+\frac{\beta}{2\alpha})\big)^2]dt.
$$
With the change of variables $\tau = e^{i\pi/4}\sqrt{\alpha} (t+\frac{\beta}{2\alpha})$, we arrive at:
$$
\mathcal{I}(\alpha,\beta)=\frac{e^{-i\pi/4}}{\sqrt{\alpha}}\exp[-i\frac{\beta^2}{4\alpha}]\int_a^be^{\tau^2}d\tau,
$$
where $a=e^{i\pi/4}\sqrt{\alpha}\beta/(2\alpha)$ and $b=e^{i\pi/4}\sqrt{\alpha}(t_f+\beta/(2\alpha))$. This integral can be computed by using the imaginary error function, Erfi:
$$
\textrm{Erfi}(x)=\frac{2}{\sqrt{\pi}}\int_0^x e^{t^2}dt.
$$
We deduce that:
$$
\mathcal{I}(\alpha,\beta)=\frac{e^{-i\pi/4}}{\sqrt{\alpha}}\exp[-i\frac{\beta^2}{4\alpha}]\frac{\sqrt{\pi}}{2}[\textrm{Erfi}(b)-\textrm{Erfi}(a)].
$$

For an ensemble of springs, the final state at time $t_f$ is given by:
$$
z_\omega(t_f)=e^{i\omega t_f}\int_0^{t_f}e^{-i\omega t}u_0\cos(\omega_i t+s\frac{t^2}{2})dt.
$$
We get
\begin{eqnarray*}
& & z_\omega(t_f)=\frac{e^{i\omega t_f}}{2}\int_0^{t_f}dt(\exp[i\frac{s}{2}t^2+i(\omega_i-\omega)t]\\
& & +\exp[-i\frac{s}{2}t^2-i(\omega_i+\omega)t])
\end{eqnarray*}
and thus
$$
z_\omega(t_f)=\frac{e^{i\omega t_f}}{2}[\mathcal{I}(\frac{s}{2},\omega_i-\omega)+\mathcal{I}(-\frac{s}{2},-\omega_i-\omega)].
$$
\section{A general Shortcut To Adiabaticity approach}\label{appb}
We describe here a general method to derive control fields based on inverse engineering approach. It can be applied to any linear control system which fulfills specific properties given below. To simplify the discussion, we assume that the system is controllable, i.e. that the Kalman criterion is satisfied~\cite{bryson,brockett,rouchon}. We consider the following linear control system:
$$
\dot{x}=Ax+Bu,
$$
where $A\in M_{n,n}(\mathbb{R})$ and $B\in M_{n,m}(\mathbb{R})$ are two constant matrices. The control term $u(t)$ is such that $u(t)\in \mathbb{R}^m$ and the state of the system $x(t)\in \mathbb{R}^n$. The goal of the control is to bring the system from $x_0$ to $x_f$ in a time $t_f$. Note that, without loss of generality, we can assume that $x_0=0$ by replacing $x_f$ by $x_f-e^{At_f}x_0$. The Kalman criterion states here that the rank of the controllability matrix $C(A,B)$ defined by:
$$
C(A,B)=[B,AB, \cdots, A^{n-1}B]
$$
is $n$. $C(A,B)$ is a $nm\times n$ matrix, where the different matrices $A^kB$ are reshaped into vectors. We also know that the set of reachable points is the image of $C(A,B)$.

We introduce a time-dependent vector $g(t)\in \mathbb{R}^m$ and the coefficients $g_k\in \mathbb{R}$ such that:
\begin{equation}\label{equ}
u(t)=\sum_{k=0}^n g_kg^{(k)}(t).
\end{equation}
We denote by $G(t_f)$ the integral:
$$
G(t_f)=\int_0^{t_f}e^{-At}Bg(t)dt.
$$
The goal of the control procedure is to find a field, $u(t)$, so that:
\begin{equation}\label{equsta}
e^{-At_f}x_f=\int_0^{t_f}e^{-At}Bu(t)dt.
\end{equation}
We consider a $g$- function with the initial conditions $g^{(0)}(0)=g^{(1)}(0)=\cdots =g^{(n-1)}(0)=0$. We have:
\begin{equation}\label{eqsyst}
\begin{cases}
\int_0^{t_f}e^{-At}Bg^{(1)}(t)dt = e^{-At_f}Bg^{(0)}(t_f)+AG(t_f) \\
\int_0^{t_f}e^{-At}Bg^{(2)}(t)dt =  e^{-At_f}(Bg^{(1)}(t_f)+ABg^{(0)}(t_f))\\
+A^2G(t_f) \\
\cdots \\
\int_0^{t_f}e^{-At}Bg^{(n)}(t)dt =  e^{-At_f}\sum_{j=0}^{n-1} A^jBg^{(n-1-j)}(t_f)\\
+A^nG(t_f)
\end{cases}
\end{equation}
Equation~\eqref{equsta} can be rewritten by plugging the expression \eqref{equ} of $u(t)$ and by using Eq.~\eqref{eqsyst}. We obtain the sum of two terms. The first one $\sum_{k=0}^ng_k A^k G(t_f)$ is equal to zero if we choose the $g_k$- coefficients as the coefficients $p_k$ of the characteristic polynomial of $A$. Indeed, from the Cayley-Hamilton theorem~\cite{brockett}, we have $\sum_{k=0}^np_k A^k=0$, where we set $p_n=1$. We finally arrive at:
\begin{eqnarray*}
x_f&=&B\sum_{k=0}^{n-1}g_{k+1}g^{(k)}(t_f)+AB\sum_{k=0}^{n-2}g_{k+2}g^{(k)}(t_f)\\
& & +\cdots +A^{n-1}Bg_ng^{(0)}(t_f).
\end{eqnarray*}
If the Kalman criterion is satisfied then any vector of $\mathbb{R}^n$, and in particular $x_f$, can be written as a sum of the form:
$$
x_f=\sum_{k=0}^{n-1}A^kB b_k
$$
where $b_k\in\mathbb{R}^m$. We obtain the following linear system:
\begin{equation}\label{eqxg}
\begin{cases}
b_{n-1}=g_ng^{(0)}(t_f) \\
b_{n-2}=g_ng^{(1)}(t_f)+g_{n-1}g^{(0)}(t_f) \\
\cdots \\
b_{0}=\sum_{k=0}^{n-1}g_{k+1}g^{(k)}(t_f)
\end{cases}
\end{equation}
Using $g_n=1$, we can deduce the final conditions $g^{(k)}(t_f)$. The $g$- function may be obtained by polynomial interpolation, but other function bases can be used. We consider a polynomial of order $2n-1$ to fulfill the $2n$ boundary conditions:
$$
g(t)=\sum_{k=n}^{2n-1}a_k(\frac{t}{t_f})^k.
$$
The first $n$ vectors $a_k\in\mathbb{R}^m$ are zero by construction and the others can be computed from the successive derivatives of $g$ by inverting the system~\eqref{eqxg}. As an illustrative example of the general approach, we consider the case of two springs and we show how to find the control field derived in Sec.~\ref{sec4} and Sec.~\ref{controltwolevel}. We have:
$$
\begin{pmatrix}
\dot{x}_1 \\
\dot{y}_1 \\
\dot{x}_2 \\
\dot{y}_2
\end{pmatrix}
=
\begin{pmatrix}
0 & -\omega_1 & 0 & 0 \\
\omega_1 & 0 & 0 & 0 \\
0 & 0 & 0 & -\omega_2 \\
0 & 0 & \omega_2 & 0
\end{pmatrix}
\begin{pmatrix}
x_1 \\
y_1 \\
x_2 \\
y_2
\end{pmatrix}
+
u(t)
\begin{pmatrix}
1 \\
0 \\
1 \\
0
\end{pmatrix}
$$
where the indices $1$ and $2$ denote respectively the first and second springs. We assume that the control field $u$ can be expressed as:
$$
u(t)=\sum_{k=0}^3g_kg^{(k)}(t)
$$
The coefficients $g_k$ are given by the characteristic polynomial $P_A$ of $A$:
$$
P_A=\lambda^4+(\omega_1^2+\omega_2^2)\lambda^2+\omega_1^2\omega_2^2,
$$
i.e. $g_4=1$, $g_2=\omega_1^2+\omega_2^2$ and $g_0=\omega_1^2\omega_2^2$, $g_3=g_1=0$. The controllability matrix $C(A,B)$ is given by the following vectors:
$$
B=\begin{pmatrix} 1 \\ 0 \\ 1 \\ 0\end{pmatrix},~AB=\begin{pmatrix} 0 \\ \omega_1 \\ 0 \\ \omega_2\end{pmatrix},
$$
and
$$
A^2B=\begin{pmatrix} -\omega_1^2 \\ 0 \\ -\omega_2^2 \\ 0\end{pmatrix},~A^3B=\begin{pmatrix} 0 \\ -\omega_1^3 \\ 0 \\ -\omega_2^3\end{pmatrix}.
$$
If $\omega_1=\pm \omega_2$ then the rank of $C(A,B)$ is strictly smaller than 4 and any target state of $\mathbb{R}^4$ cannot be reached. For a robust control of two springs, the target state is $x_f=(1,0,1,0)^\intercal$. We deduce that $b_0=1$, $b_1=0$, $b_2=0$ and $b_3=0$ and the corresponding boundary conditions for $g$. It is then straightforward to derive the $g$- function of Eq.~\eqref{eqg4} and to extend this computation to $N$ frequencies for the $g$- function of Eq.~\eqref{eqgN}.
The target state for the selective control of Sec.~\ref{controltwolevel} is $x_f=(\frac{\pi}{2},0,0,0)^\intercal$. The first step consists in solving the equation:
$$
x_f=Bb_0+ABb_1+A^2Bb_2+A^3Bb_3
$$
For the frequencies $\omega_1=0$ and $\omega_2=0.5$, we get $(b_0,b_1,b_2,b_3)=(\frac{\pi}{2},0,2\pi,0)$. Using Eq.~\eqref{eqxg}, we obtain the final boundary conditions for the $g$- function, $(g^{(3)}(t_f),g^{(2)}(t_f),g^{(1)}(t_f),g^{(0)}(t_f))=(0,0,2\pi,0)$, and then the coefficients of the polynomial.
\\ \\
\noindent\textbf{ACKNOWLEDGMENT}\\
D. Sugny thanks K. Beauchard and M. A. Delsuc for helpful discussions. D. Sugny acknowledges support from the QUACO project (ANR 17-CE40-0007-01). This project has received funding from the European Union's Horizon 2020 research and innovation programme under the Marie-Sklodowska-Curie grant agreement No 765267 (QUSCO). This work has been financially supported by the Agence Nationale de la Recherche research funding Grant No. ANR-18-CE30-0013.


\begin{thebibliography}{50}

\bibitem{bryson} A. E. Bryson and Y.-C. Ho, \emph{Applied optimal control} (Taylor \& Francis, New York, 1975).

\bibitem{bressan} A. Bressan and B. Piccoli, \emph{Introduction to the Mathematical Theory of Control} (American Institute of Mathematical Sciences, Springfield, 2007).

\bibitem{glaserreview} S. J. Glaser, U. Boscain, T. Calarco, C. Koch, W. Kockenberger, R. Kosloff, I. Kuprov, B. Luy, S. Schirmer, T. Schulte-Herbr\"uggen, D. Sugny and F. Wilhelm, Eur. Phys. J. D {\bf 69}, 279 (2015)

\bibitem{schattler} H. Sch\"attler and U. Ledzewicz, \emph{Geometric optimal control: Theory, Methods and Examples} (Springer, New York, 2010)

\bibitem{brifreview} C. Brif, R. Chakrabarti and H. Rabitz, New J. Phys. {\bf 12}, 075008 (2010)

\bibitem{RMP19} C. P. Koch, M. Lemeshko and D. Sugny, Rev. Mod. Phys. {\bf 91}, 035005 (2019)

\bibitem{dongreview} D. Dong and I. A. Petersen, IET Control Theory A. {\bf 4}, 2651 (2010)

\bibitem{altafinireview} C. Altafini and F. Ticozzi, IEEE Trans. Automat. Control {\bf 57}, 1898 (2012)

\bibitem{adiabaticreview} N. V. Vitanov, T. Halfmann, B. W. Shore, and K. Bergmann, Annual review of physical chemistry {\bf 52},
763 (2001).

\bibitem{stirapRMP} N. V. Vitanov, A. A. Rangelov, B. W. Shore, K. Bergmann, Rev. Mod. Phys. {\bf 89}, 015006 (2017)

\bibitem{pont} L. S. Pontryagin et al., \emph{The Mathematical Theory of Optimal Processes} (John Wiley and Sons, New York,
London, 1962)

\bibitem{bonnardbook} B. Bonnard and D. Sugny, \emph{Optimal control in space and quantum dynamics} (AIMS applied Math. Vol. 5, 2012)

\bibitem{jurdjevicbook} V. Jurdjevic, \emph{Geometric control theory} (Cambridge University Press, Cambridge, 1996)

\bibitem{garon} A. Garon, S. J. Glaser and D. Sugny, Phys. Rev. A {\bf 88}, 043422 (2013)

\bibitem{reviewSTA1} E. Torrontegui, S. Ib\'anez, S. Mart\'inez-Garaot, M. Modugno, A. del Campo, D. Gu\'ery-Odelin, A. Ruschhaupt, X. Chen and J. Gonzalo Muga,  Adv. At. Mol. Opt. Phys. \textbf{62}, 117 (2013).

\bibitem{reviewSTA2} S. Deffner, C. Jarzynski, and A. del Campo, Phys. Rev. X \textbf{4}, 021013 (2014).

\bibitem{reviewSTA3} M. Kolodrubetz, D. Sels, P. Mehta, and A. Polkovnikov, Physics Reports \textbf{697}, 1 (2017).

\bibitem{reviewSTA4} D. Gu\'ery-Odelin, A. Ruschhaupt, A. Kiely, E. Torrontegui, S. Mart\'inez-Garaot, J. G. Muga, Rev. Mod. Phys. \textbf{91}, 045001 (2019).

\bibitem{liens1} J. S. Li and N. Khaneja, Phys. Rev. A \textbf{73}, 030302 (2006)

\bibitem{liens2} J. S. Li and N. Khaneja, IEEE Trans. Autom. Control \textbf{54}, 528 (2009)

\bibitem{kozbar2012} K. Kozbar, S. Ehni, T. E. Skinner, S. J. Glaser, and B. Luy, J. Magn. Reson. \textbf{225}, 142 (2012)

\bibitem{kozbar2004} K. Kobzar, T. E. Skinner, N. Khaneja, S. J. Glaser, and B. Luy, J. Magn. Reson. \textbf{170}, 236 (2004)

\bibitem{skinner2012} T. E. Skinner, N. I. Gershenzon, M. Nimbalkar, W. Bermel, B. Luy, and S. J. Glaser, J. Magn. Reson. \textbf{216}, 78 (2012).

\bibitem{vandamme2017} L. Van Damme, S. J. Glaser and D. Sugny, Phys. Rev. A \textbf{95}, 063403 (2017)

\bibitem{STAnjp} A. Ruschhaupt, X. Chen, D. Alonso and J. G. Muga, New J. Phys. {\bf 14}, 093040 (2012).

\bibitem{daemsprl} D. Daems, A. Ruschhaupt, D. Sugny and S. Gu\'erin, Phys. Rev. Lett. {\bf 111}, 050404 (2013).

\bibitem{vandamme} L. Van Damme, D. Schraft, G. Genov, D. Sugny, T. Halfmann and S. Gu\'erin, Phys. Rev. A {\bf 96}, 022309 (2017)

\bibitem{grape} N. Khaneja, T. Reiss, C. Kehlet, T. Schulte-Herbr\"{u}ggen and S. J. Glaser, J. Magn. Reson. {\bf 172}, 296 (2005)

\bibitem{reichkrotov} D. M. Reich, M. Ndong and C. P. Koch, J. Chem. Phys. {\bf 136}, 104103 (2012)

\bibitem{gross} J. Werschnik and E. K. U. Gross, J. Phys. B {\bf 40}, R175 (2007)

\bibitem{liberzon} D. Liberzon, \emph{Calculus of variations and Optimal control theory} (Princeton University Press, Princeton, 2012).

\bibitem{brockett} R. W. Brockett, \emph{Finite Dimensional Linear Systems}, (John Wiley and Sons, New York, 1970).

\bibitem{rouchon} F. Bonnans and P. Rouchon, \emph{Commande et optimisation de syst\`emes dynamiques}, (Ecole Polytechnique, Paris, 2006).

\bibitem{Lithesis} J.-S. Li, \emph{Control of inhomogeneous ensemble}, PhD thesis in Applied Mathematics, Harvard University (2006).

\bibitem{vartan} V. Martikyan, D. Gu\'ery-Odelin and D. Sugny, Phys. Rev. A \textbf{101}, 013423 (2020)

\bibitem{li:2017} J.-S. Li, J. Ruths and S. J. Glaser, Nat. Comm. {\bf 8}, 446 (2017).



\bibitem{li:2011} J.-S. Li, IEEE Trans. A. C. \textbf{56}, 345 (2011).

\bibitem{lilin} A. Zlotnik and J. S. Li, American Control Conference, Montreal, 5849 (2012), doi: 10.1109/ACC.2012.6315297.

\bibitem{guery:2014} D. Gu\'ery-Odelin and J. G. Muga, Phys. Rev. A {\bf 90}, 063425 (2014)

\bibitem{bodenhausen:2016} A. A. Sehgal, P. Pelupessy, C. Rolando and G. Bodenhausen, Chem. Phys. \textbf{18}, 9167 (2016).

\bibitem{delsuc:2013} M. A. Van Agthoven, M.-A. Delsuc, G. Bodenhausen and C. Rolando, Anal. Bioanal. Chem. \textbf{405}, 51 (2013).

\bibitem{delsuc:2017} F. Bray, J. Bouclon, L. Chiron, M. Witt, M.-A. Delsuc and C. Rolando, Anal. Chem. \textbf{89}, 8589 (2017).

\bibitem{delsuc:2016} F. Floris, M. van Agthoven, L. Chiron, A. J. Soulby, C. A. Wootton, Y. P. Y. Lam, M. P. Barrow, M.-A. Delsuc, P. B. O'Connor, J. Am. Soc. Mass Spectrom. \textbf{27}, 1531 (2016).

\bibitem{slope1} N. I. Gershenzon, T. E. Skinner, B. Brutscher, N. Khaneja, M. Nimbalkar, B. Luy and S. J. Glaser, J. Magn. Reson. \textbf{192}, 235 (2008)

\bibitem{slope2} N.I. Gershenzon, K. Kobzar, B. Luy, S.J. Glaser, T.E. Skinner, J. Magn. Reson. \textbf{188}, 330 (2007)

\bibitem{genov:2014} G. T. Genov, D. Schraft, T. Halfmann, and N. V. Vitanov, Phys. Rev. Lett. \textbf{113}, 043001 (2014).

\bibitem{levitt:1986} M. H. Levitt, Prog. Nucl. Magn. Reson. Spectrosc. \textbf{18}, 61 (1986).

\bibitem{rong:2015} X. Rong, J. Geng, F. Shi, Y. Liu, K. Xu, W. Ma, F. Kong, Z. Jiang, Y. Wu, and J. Du, Nat. Commun. \textbf{6}, 8748 (2015).

\bibitem{selective} L. Van Damme, Q. Ansel, S. J. Glaser and D. Sugny, Phys. Rev. A \textbf{98}, 043421 (2018)

\bibitem{levittbook} M. H. Levitt, \emph{Spin Dynamics: Basics of Nuclear Magnetic Resonance} (Wiley, New York, 2008).

\bibitem{lapert:2010} M. Lapert, Y. Zhang, M. Braun, S. J. Glaser, and D. Sugny, Phys. Rev. Lett. \textbf{104}, 083001 (2010).


%\bibitem{stefanatos:2009} J.-S. Li, J. Ruths, D. Stefanatos, J. Chem. Phys. {\bf 131}, 164110 (2009)

%\bibitem{alessandro} D. D'Alessandro, IEEE Trans. Autom. Control {\bf 46}, 866 (2001)

%\bibitem{boscain} U. Boscain and P. Mason, J. Math. Phys. {\bf 47}, 062101 (2006)



%\bibitem{khanejaspin} N. Khaneja, R. Brockett and S. J. Glaser, Phys. Rev. A {\bf 63}, 032308 (2001)

%\bibitem{bonnard:2012} B. Bonnard, O. Cots, S. J. Glaser, M. Lapert, D. Sugny and Y. Zhang, IEEE transactions on automatic and control, \textbf{57}, 8, 1957 (2012)

%\bibitem{lapert:2012} M. Lapert, G. Ferrini and D. Sugny, Phys. Rev. A \textbf{85}, 023611 (2012)














%\bibitem{stefanatos:2010} D. Stefanatos, J. Ruths and J. S. Li, Phys. Rev. A {\bf 82}, 063422 (2010)

%\bibitem{xen:2011} X. Chen, E. Torrontegui, D. Stefanatos, J.-S. Li and J. G. Muga, Phys. Rev. A {\bf 84}, 043415 (2011).





%\bibitem{motionplanning} S. M. LaValle, \emph{Planning Algorithms} (Cambridge University Press, Cambridge, U.K., 2006).








\end{thebibliography}
\end{document}